\documentclass[12pt,prd,onecolumn,floatfix,amsmath,nofootinbib,amssymb,floatfix]{revtex4}
\usepackage{amssymb}
\usepackage{amsmath}
\usepackage{graphicx,color,dcolumn,booktabs,bm}
\usepackage{longtable,lscape}
\usepackage{txfonts}
\usepackage{overpic}
\usepackage{amssymb}
\usepackage{indentfirst}
\usepackage{feynmf}   %{feynmp}
\usepackage{slashed}  %for Feynman symbols
\usepackage{cases}
\usepackage{color}
\usepackage{multirow}
\usepackage{slashed}
\usepackage{epstopdf}
\usepackage{longtable}
\usepackage{ulem}
\usepackage{graphicx,color,dcolumn,booktabs,bm}
\usepackage[colorlinks,
            citecolor=blue,
            anchorcolor=red,
            menucolor=red,
            linkcolor=red,
            filecolor=red,
            runcolor=red,
            urlcolor=blue,
            frenchlinks=red]{hyperref}
\graphicspath{{Figures/}} %
\allowdisplaybreaks

\begin{document}
\title{A Mixing Coupling Scheme for Spectra of Singly Heavy Baryons with Spin-1 Diquarks in P-waves}

\author{Duojie Jia$^{1,4}$
\footnote{Corresponding author}}
\email{jiadj@nwnu.edu.cn}

\author{Ji-Hai Pan$^{1}$}
\author{Cheng-Qun Pang$^{2,3,4}$}\email{pcq@qhnu.edu.cn}

\affiliation{$^1$Institute of Theoretical Physics, College of Physics and
Electronic Engineering, Northwest Normal University, Lanzhou 730070, China
\\
$^2$College of Physics and Electronic Information Engineering,
Qinghai Normal University, Xining 810000, China\\
$^3$Joint Research Center for Physics, Lanzhou University
and Qinghai Normal University, Xining 810000,China \\ 
$^4$Lanzhou Center for Theoretical Physics, Lanzhou University,
Lanzhou 730000,China \\}

\begin{abstract}
A new scheme of state classification is proposed and applied to
analyze masses of the heavy baryons $\Omega _{Q}$, $\Sigma _{Q}$ and $\Xi_{Q}^{\prime}$ in P-waves. The results confirm all excited $\Omega _{c}$ and $\Omega _{b}$ baryons reported recently by LHCb to be bound states of a P-wave $ss$-diquark and a respective charm or bottom quark, and thereby predict Regge trajectories for more excited $\Omega _{c}$ and $\Omega _{b}$ baryons. We suggest one excited {$J^{P}=5/2^{-}$} $\Omega _{b}$ state to be unseen by LHCb around $6352$ MeV, and predict P-wave masses of all spin-partners of the odd-parity baryons $\Sigma_{c}(2800)/\Xi_{c}^{\prime}(2942)$ and
$\Sigma_{b}(6097)$/$\Xi _{b}^{\prime }(6227)$. A computation is further given in a relativized potential quark models to explain matched values of spin couplings of all considered baryons, by which a scaling law for these spin couplings is discussed.

{\normalsize PACS number(s):12.39Jh, 12.40.Yx, 12.40.Nn}

\end{abstract}
\maketitle
\date{\today}
\maketitle\section{Introduction}
Scientific interest as to spectrum of strongly interacting heavy
hadrons has raised again in last two decades greatly due to the discoveries of
numerous heavy subatomic particles \cite{Tanabashi:D18}. In 2017, LHCb
observed five even narrower $\Omega _{c}$ resonances \cite{Aaij:L2017}(Table I) in the $\Xi _{c}^{+}K^{-}$ decay channel based on $pp$ collision data. Very recently, LHCb, again, reported the discovery of four even narrower $\Omega
_{b}$ resonances around $6.3$ GeV decaying into $\Xi _{b}^{0}K^{-}$ \cite{Aaij:L2020}, with slightly less significance. The reported masses and decay widths are listed in Table I for both of them.

Observed patterns of masses of the excited $\Omega _{c,b}$ baryons turn out
to be perplexing and thereby rise naturally questions as to (i) why there are four
excited states of the $\Omega _{b}$ baryons while there are five for the $%
\Omega _{c}$, and (ii) what their spin-parities or their inner structures
are. These questions have been addressed by many authors  \cite%
{EFG:B08,EFG:D11,MaltmanI:D80,RobertsP:A08,Garcilazo:G07,Migura:A06,Valcarce:A08,YamaguchiH:D15,PerezCB:D15,ShahTh:C16,YoshidaHH:D15,ChenHL:D15,ChenMH:D16}
with recent explorations given in Refs  \cite%
{KarlinerR:17,PadmanathM:L17,ChenL:D17,WangZ:D17,Wang:C17,Agaev:U17,ChenHL:D17,ChengC:D17,WangZZ:D17,ZhaoY:D17,YangP:D18,HuangP:D18,AnChen:D17,Prasz:D17,KimPP:D17,ChenLM:E17,KarlinerR:Ax20}%
, where spin-parities were assigned for the $\Omega _{c}$ states around $3.0\pm0.1$GeV and for the excited $\Omega _{b}$ states around $6.3\pm 0.2$GeV. Regarding the reported $\Omega _{c}$ resonances, different
 interpretations were suggested, which include the $P$-wave assignment \cite%
{KarlinerR:17,PadmanathM:L17,ChenL:D17,WangZ:D17,Wang:C17,ShahTh:C16}, the $P
$-wave and $S/D$-wave assignment \cite{ChengC:D17,ZhaoY:D17,ChenHL:D15}, the
charmed exotic systems \cite{YangP:D18,HuangP:D18,AnChen:D17,Prasz:D17,KimPP:D17}, and entail more efforts to answer above two questions.

For charmed baryons, spectroscopy is known to be intricate, with many
excited states expected \cite%
{EFG:B08,EFG:D11,MaltmanI:D80,RobertsP:A08,Garcilazo:G07,PerezCB:D15,YoshidaHH:D15,ChenHL:D15}%
. The common practice is to utilize the heavy quark
symmetry(HQS) \cite{Manohar:D07}, which becomes exact in the limit of heavy
quark $m_{Q}\rightarrow \infty $. Assuming HQS by which spin $\mathbf{S}_{Q}$
of heavy quark $Q$ is conserved, one can classify heavy hadrons by their
total spin $\mathbf{J}$ and the angular momentum $\mathbf{j=J-S}_{Q}$ of
light degrees of freedom(named $jj$ coupling). This is analogous to the
hydrogen-like atoms whose quantum states are well labeled by quantum numbers
of outermost electron. While this picture works well in describing
the normal charmed and bottom baryons \cite{EFG:D11,MaltmanI:D80,RobertsP:A08}
with strangeness $|S|<2$, its applicability to the $\Omega _{c,b}$ baryons
remains to be explored.

The purpose of this work is to explore quantum numbers and inner structures of the odd-parity bound systems of a heavy $Q$-quark ($Q=c,b$) and a light spin-1 $qq$-diquark($qq=nn,ns, n=u,d$) and corresponding main feature of relative magnitudes of the spin coupling parameters for the systems. We propose a new scheme of state classification (named the $Jls$ mixing coupling) to perform systematic mass analysis for the newly LHCb-reported resonances of the excited $\Omega_{c}$ and $\Omega_{b}$, which interprets all of these resonances to be the $P$-wave negative-parity baryons, and enables us to predict Regge trajectories of the mean-spin masses of their excited states. We update earlier computations for the negative-parity baryons $\Sigma_{Q}$ and $\Xi _{Q}^{\prime }$, and explain the matched spin-couplings of these baryons in a relativized potential quark model,

\smallskip

\begin{table}[htb]
\caption{Masses and widths of the LHCb-reported resonances $\Omega _{c}$ \cite{Aaij:L2017} and $\Omega_{b}$\cite{Aaij:L2020}, and excited baryons $\Sigma _{Q}$ and $\Xi_{Q}^{\prime}$\cite{Tanabashi:D18}, with the proposed values of spin-parity $J^{P}$ shown also.   \label{Table 1}}
\[\begin{array}{cccc}
\hline\hline
$State$               &$Mass (MeV)$            & $Width (MeV)$             &$$J^{P}$ Proposed$ \\\hline
\Omega _{c}(3000)^{0} &3000.4\pm 0.2\pm 0.1   &4.5\pm 0.6\pm 0.3         &1/2^{-} \\
\Omega _{c}(3050)^{0} &3050.2\pm 0.1\pm 0.1  & 0.8\pm 0.2\pm 0.1^{}\footnote{
In total, it is $<1.2$ MeV, $95\%$ C.L. $^{b}$It is $<1.2$ MeV, $95\%$ C.L
totally.}  &1/2^{-} \\
\Omega _{c}(3066)^{0} &3065.6\pm 0.1\pm 0.3  & 3.5\pm 0.4\pm 0.2      &3/2^{-} \\
\Omega _{c}(3090)^{0} &3090.2\pm 0.3\pm 0.5  & 8.7\pm 1.0\pm 0.8      &3/2^{-} \\
\Omega _{c}(3119)^{0} &3119.1\pm 0.3\pm 0.9  & 1.1\pm 0.8\pm 0.4^{b}  &5/2^{-} \\
\Omega _{b}(6316)^{-} &6315.64\pm 0.31\pm 0.07^{\pm 0.5} & <2.8,     &1/2^{-} \\
\Omega _{b}(6330)^{-} &6330.30\pm 0.28\pm 0.07^{\pm 0.5} & <3.1,     &1/2^{-} \\
\Omega _{b}(6340)^{-} &6339.71\pm 0.26\pm 0.05^{\pm 0.5} & <1.5      &3/2^{-} \\
\Omega _{b}(6350)^{-} &6349.88\pm 0.35\pm 0.05^{\pm 0.5} & 1.4_{-0.8}^{+0.1}\pm 1.0 & 3/2^{-} \\
\Sigma_{c}(2800)^{++} &2801_{-6}^{+4}  & 75_{-17}^{+22}    &3/2^{-} \\
\Xi_{c}^{\prime}(2930) &2942\pm 5      & 36  \pm 13        & 3/2^{-} \\
\Sigma_{b}(6097)^{-}   &6098.0 \pm 1.8 & 29\pm  4          & 3/2^{-} \\
\Xi_{b}^{\prime}(6227)^{-} & 6226.9\pm2.0  &  18\pm 6      & 1/2^{-} \\
\hline\hline
\end{array}\]
\end{table}

To achieve that, we determine the mean-spin mass $\bar{M}$ and the spin coupling strengths ($a_{1},a_{2},b,c$) as effective parameters by matching computed and measured masses of heavy baryons considered, which implies that a central potential plus spin-interactions describe the main features of most established hadrons.  Due to their connections with the interquark potential, these knowledges about spin-couplings are useful and forms a basis for further understanding QCD interaction within hadrons.

We apply our proposed scheme of the $Jls$ mixing coupling to mass analysis of the excited $\Omega _{c}$ baryons in Sect. II, where preferred spin-parity are assigned for them. We examine other plausible spin-parity arrangements in Sect. III. Similar mass analysis is given for excited $\Omega _{b}$ baryons in Sect. IV. In Sect. V, the $Jls$ mixing coupling is applied to the less strange baryons $\Sigma_{Q}$ and $\Xi _{Q}^{\prime }$. A computation is given in the relativized quark model to explain the matched spin-couplings in Sect. VI. We end with conclusions and discussions in Sect. VII.

\section{Inner Structures of the $c(ss)$ Systems}

In heavy quark-diquark picture, two strange quarks in a $c(ss)$
system form a S-wave anti-color triplet($\mathbf{\bar{3}}_{c}$) diquark $%
(ss) $, with spin one($S_{ss}=1$) due to the spatial symmetry under exchange
of two fermions. The diquark spin($=1$) can couple with spin $S_{c}=1/2$ of
the charm quark $c$ to form a total spin $S=1\pm 1/2=1/2,3/2$. Let us
consider the relative P-wave excitations of diquark $ss$ with respect to the
charm quark $c$(the relative orbital angular momentum $L=1$). Coupling of $%
L=1$ with the spin $S=1/2$ gives states with total spin $J=1/2,3/2$, while
coupling with $S=3/2$ leads to states with $J=1/2,3/2$ and $5/2$. Symbolically, they are written as, in terms of adding of angular momentum,
\begin{eqnarray}
\left( \frac{1}{2}\right) _{S}\otimes 1_{L} &=&\frac{1}{2}\oplus \frac{3}{2},
\notag \\
\left( \frac{3}{2}\right) _{S}\otimes 1_{L} &=&\frac{1}{2}^{\prime }\oplus
\frac{3}{2}^{\prime }\oplus \frac{5}{2},  \label{LSc}
\end{eqnarray}%
In total, one has five $P$-wave states with $%
J=1/2,3/2,1/2^{^{\prime }}$, $3/2^{\prime }$ and $5/2$ with negative parity $%
P=-1$ which implies $5!=120$ a priori possible assignments for their $J^{P}$
quantum numbers.

For most heavy baryons with less strangeness, the $jj$ coupling has been
commonly used to classify their states in terms of hadron eigenstates $%
|J,j\rangle $, with $J$ the eigenvalues of $\mathbf{J}$ and $j$ of $\mathbf{j%
}$ respectively. In the case of doubly strange $Qss$ baryons with the $ss$-diquark comparable with the heavy quark $Q$ in mass, the finite mass effect of the heavy quark may become important and makes it appropriate to go beyond the $jj$ coupling.

For this, we propose a new scheme of state classification(named the $Jls$
mixing coupling) in which the eigenfunctions (bases) of spin multiplets of
the $Q(ss)$ system($Q=c,b$) diagonalize all spin-orbit interactions between
the heavy quark $Q$ and the diquark $ss$. The spin-dependent interaction for
the $Q(ss)$ systems, in the heavy quark-diquark picture, is \cite%
{EFG:D11,KarlinerR:D15}
\begin{equation}
H^{SD}=a_{1}\mathbf{L}\cdot \mathbf{S}_{ss}+a_{2}\mathbf{L}\cdot \mathbf{S}%
_{Q}+bS_{12}+c\mathbf{S}_{ss}\cdot \mathbf{S}_{Q},  \label{Hsd}
\end{equation}%
\begin{equation*}
S_{12}=3\mathbf{S}_{ss}\cdot \mathbf{\hat{r}S}_{Q}\cdot \mathbf{\hat{r}}-%
\mathbf{S}_{ss}\cdot \mathbf{S}_{Q},
\end{equation*}%
where the first two terms are spin-orbit interactions, the third is the
tensor energy, and the last the contact interaction between the heavy quark
spin $\mathbf{S}_{Q}$ and the diquark spin $\mathbf{S}_{ss}$. Here, $\mathbf{%
L}$ is the orbital angular momentum of the system. Though four strengths $a_{1,2}$, $b$ and $c$ of spin-couplings will be treated, in a sense, as effective parameters in this work, they are closely related to interquark interactions within hadrons and assumed, in potential quark model\cite{EFG:D11,MaltmanI:D80,RobertsP:A08,YoshidaHH:D15,ChenLM:E17}, to be computable from QCD analogues of Breit-Fermi interaction in QED.

If $Q$ is very heavy, the first spin-orbit term $\mathbf{L}\cdot \mathbf{S}%
_{ss}$ should dominate over the others in Eq.(\ref{Hsd}) if one assumes
spin-dependent interactions enter the color hyperfine interactions through
the magnetic moment $e_{i}\mathbf{S}_{i}/m_{i}$ of the quark $i=Q$ or $ss$
with mass $m_{i}$, by analogy with the spin-relevant relativistic correction
of the heavy quarkonium \cite{Yndurain83}. As such, the heavy quark spin $%
\mathbf{S}_{Q}$ decouples with the light degree of freedom and is conserved
in the heavy quark limit, making the basis eigenfunctions of $\mathbf{L}%
\cdot \mathbf{S}_{ss}$(the $jj$ coupling) appropriate to classify the baryon
states. Here, the baryon spin $\mathbf{J}$ and the diquark angular momentum $%
\mathbf{j=J-S}_{Q}$ are both conserved, enabling baryon states to be labelled by
the $jj$ coupling states $|J,j\rangle $. In the case of the $c(ss)$ or $b(ss)$
baryons, in which the diquark mass $m_{ss}$(near $1$ GeV) is comparable to $M_{Q}$(about $1.5$ GeV), the terms like $a_{2}\mathbf{L}\cdot \mathbf{S}_{Q}$
in Eq. (\ref{Hsd}) may become important. Taking $a_{1}=a_{2}$, for instance, the
spin-orbit interaction becomes proportional to $\mathbf{L}\cdot \lbrack
\mathbf{S}_{ss}+\mathbf{S}_{Q}]=\mathbf{L}\cdot \mathbf{S}$ and diagonal in
the $LS$ coupling, in which $\mathbf{S}_{Q}$ first couples $\mathbf{S}_{ss}$ to
form the total quark spin $\mathbf{S}=\mathbf{S}_{ss}+\mathbf{S}_{Q}$ and then to $\mathbf{L}$ of the quark-diquark system to form $\mathbf{J}$,
as shown in Eq. (\ref{LSc}).

In the $Jls$ mixing coupling considered in this work, the bases $|J,j_{LS}=j^{\prime }\rangle $ diagonalize the interaction $a_{1}\mathbf{L}\cdot \mathbf{S}_{ss}+a_{2}\mathbf{L}\cdot \mathbf{S}_{Q}+bS_{12}$, instead of $\mathbf{L}\cdot \mathbf{S}_{ss}$ solely, in Eq.(\ref{Hsd}). The $Jls$ scheme reduces to the $jj$
coupling when the ratio $\epsilon =a_{2}/a_{1}$(expected to scale as $%
m_{ss}/M_{Q}$) tends to zero in the heavy quark limit,
\begin{equation}
a_{1}\mathbf{L}\cdot \lbrack \mathbf{S}_{ss}+\epsilon \mathbf{S}%
_{Q}]+b S_{12}\simeq a_{1}\mathbf{L}\cdot \mathbf{S}_{ss}\text{, as }%
M_{Q}\rightarrow \infty \text{,}  \label{tends}
\end{equation}%
where $b$ is expected to be suppressed by $1/M_{Q}$ \cite%
{KarlinerR:D15,JiaLH:19}. Finding the $Jls$ mixing eigenstates $%
|J,j_{LS}=j^{\prime }\rangle $ can be done by solving the linear eigenstate
equation of $2\times 2$ matrices $\Delta \mathcal{M}_{J}$ of the mass shift
interaction (\ref{Hsd}) in the spin subspace of $J=1/2$ and $J=3/2$ \cite%
{KarlinerR:D15,JiaLH:19}(see Ref.  \cite{Landau} also). In terms of the $LS\,$%
bases $^{2S+1}P_{J}=%
\{^{2}P_{1/2},^{4}P_{1/2},^{2}P_{3/2},^{4}P_{3/2},^{4}P_{5/2}\}$, the matrix
forms of these mass shift interactions are(see appendix A of Ref.  \cite%
{KarlinerR:17})
\begin{eqnarray}
\Delta \mathcal{M}_{J=1/2} &=&\left[
\begin{array}{cc}
\frac{1}{3}(a_{2}-4a_{1}) & \frac{\sqrt{2}}{3}(a_{2}-a_{1})+\frac{b}{\sqrt{2}%
} \\
\frac{\sqrt{2}}{3}(a_{2}-a_{1})+\frac{b}{\sqrt{2}} & -\frac{5}{3}(a_{1}+%
\frac{1}{2}a_{2})-b%
\end{array}%
\right]   \notag \\
&&+\left[
\begin{array}{cc}
-c & 0 \\
0 & \frac{1}{2}c%
\end{array}%
\right] ,  \label{M2}
\end{eqnarray}%
\begin{eqnarray}
\Delta \mathcal{M}_{J=3/2} &=&\left[
\begin{array}{cc}
\frac{2}{3}a_{1}-\frac{1}{6}a_{2} & \frac{\sqrt{5}}{3}(a_{2}-a_{1})-\frac{b}{%
2\sqrt{5}} \\
\frac{\sqrt{5}}{3}(a_{2}-a_{1})-\frac{b}{2\sqrt{5}} & -\frac{1}{3}%
(2a_{1}+a_{2})+\frac{4b}{5}%
\end{array}%
\right]   \notag \\
&&+\left[
\begin{array}{cc}
-c & 0 \\
0 & \frac{1}{2}c%
\end{array}%
\right] ,  \label{M3}
\end{eqnarray}%
\begin{equation}
\Delta \mathcal{M}_{J=5/2}=a_{1}+\frac{1}{2}a_{2}-\frac{b}{5}+\frac{c}{2}.
\label{M5}
\end{equation}

Given $J^{P}$ assignments of the five $\Omega _{c}$ states(two states of $%
J^{P}=1/2^{-}$, two of $J^{P}=3/2^{-}$, and one of $J^{P}=5/2^{-}$) there
should, in principle, exist one unique solution for the four parameters $%
a_{1}$, $a_{2}$, $b$ and $c$. As we shall find below, there is one solution
in which all states are $P$-waves with reasonable values of parameters and the mass pattern as reported by LHCb for the excited $\Omega _{c}
$ states in Table I, where our preferred $J^{P}$ assignments are shown.

Diagonalizing the mass shift operator $\mathbf{L}\cdot \mathbf{S}%
_{ss}+\epsilon \mathbf{L}\cdot \mathbf{S}_{Q}+b_{1}S_{12}$, with $b_{1}\equiv b/a_{1}$, one can compute the mass shifts $\Delta M$, the
eigenvalues of Eqs. (\ref{M2})-(\ref{M5}), by treating the contact term $c%
\mathbf{S}_{ss}\cdot \mathbf{S}_{Q}$ as a perturbation($c_{1}=c/a_{1}$ is
expected to be small for the relative $1P$-wave between diquark and $Q$).
The lowest order perturbation theory gives(see appendix A)
\begin{eqnarray}
\Delta M(J &=&1/2,0^{\prime })=-\frac{a_{1}}{4}\left( 6+\sqrt{\Delta
_{1}\left( \frac{a_{2}}{a_{1}},\frac{b}{a_{1}}\right)}+\frac{a_{2}}{a_{1}}\right) -\frac{b}{2}+c\Delta
_{3}^{+}\left( \frac{a_{2}}{a_{1}},\frac{b}{a_{1}}\right),  \notag \\
\Delta M(J &=&1/2,1^{\prime })=-\frac{a_{1}}{4}\left( 6-\sqrt{\Delta
_{1}\left( \frac{a_{2}}{a_{1}},\frac{b}{a_{1}}\right)}+\frac{a_{2}}{a_{1}} \right) -\frac{b}{2}+c\Delta
_{3}^{-}\left( \frac{a_{2}}{a_{1}},\frac{b}{a_{1}}\right),  \notag \\
\Delta M(J &=&3/2,1^{\prime })=-a_{1}\left( \sqrt{\Delta _{2}\left( \frac{a_{2}}{a_{1}},\frac{b}{a_{1}}\right)}+\frac{a_{2}}{4a_{1}}\right) +\frac{2b}{5}+c\Delta
_{4}^{+}\left( \frac{a_{2}}{a_{1}},\frac{b}{a_{1}}\right),  \notag \\
\Delta M(J &=&3/2,2^{\prime })=a_{1}\left( \sqrt{\Delta _{2}\left( \frac{a_{2}}{a_{1}},\frac{b}{a_{1}}\right)}-\frac{a_{2}}{4a_{1}}\right) +\frac{2b}{5}+c\Delta
_{4}^{-}\left( \frac{a_{2}}{a_{1}},\frac{b}{a_{1}}\right),  \notag \\
\Delta M(J &=&5/2,2^{\prime })=a_{1}+\frac{a_{2}}{2} -%
\frac{b}{5}+\frac{c}{2},  \label{MM}
\end{eqnarray}%
where six functions $\Delta _{1,2}(\epsilon ,x),\Delta _{3}^{\pm }(\epsilon
,x)$ and $\Delta _{4}^{\pm }(\epsilon ,x)$ are defined by
\begin{equation}
\begin{array}{c}
\Delta _{1}(\epsilon ,x)=4+12x^{2}+4x\left( 5\epsilon -2\right) -4\epsilon
+9\epsilon ^{2}, \\
\Delta _{2}(\epsilon ,x)=1+\frac{1}{5}x^{2}-\frac{x}{5}(1+2\epsilon
)-\epsilon +\frac{9}{16}\epsilon ^{2}.%
\end{array}
\label{del12}
\end{equation}%
\begin{equation}
\begin{array}{c}
\Delta _{3}^{+}(\epsilon ,x)=\frac{4-(2+6x+7\epsilon -3\sqrt{\Delta
_{1}(\epsilon ,x)})^{2}/(2\epsilon -2+3x)^{2}}{8+(2+6x+7\epsilon -3\sqrt{%
\Delta _{1}(\epsilon ,x)})^{2}/(2\epsilon -2+3x)^{2}}, \\
\Delta _{3}^{-}(\epsilon ,x)=\Delta _{3}^{+}\left( \sqrt{\Delta _{1}}%
\rightarrow -\sqrt{\Delta _{1}}\right) .%
\end{array}
\label{del3}
\end{equation}%
\begin{equation}
\begin{array}{c}
\Delta _{4}^{+}(\epsilon ,x)=\frac{10-(40-24x+5\epsilon +60\sqrt{\Delta
_{2}(\epsilon ,x)})^{2}/(10-10\epsilon +3x)^{2}}{20+(40-24x+5\epsilon +60%
\sqrt{\Delta _{2}(\epsilon ,x)})^{2}/(10-10\epsilon +3x)^{2}}, \\
\Delta _{4}^{-}(\epsilon ,x)=\Delta _{4}^{+}\left( \sqrt{\Delta _{2}}%
\rightarrow -\sqrt{\Delta _{2}}\right) ,%
\end{array}
\label{del4}
\end{equation}%
with $\Delta _{3,4}^{-}(\epsilon ,x)$ obtained from $\Delta
_{3,4}^{+}(\epsilon ,x)$ by merely replacing $\sqrt{\Delta _{1,2}}%
\rightarrow -\sqrt{\Delta _{1,2}}$. This expresses the mass shifts in terms
of four parameters ($a_{1},a_{2},b,c$) nonlinearly. Expanding Eqs. (\ref{MM}) to the leading order of $\epsilon ,b_{1}$ and $c_{1}$, it reduces to the linear mass shift in Ref.  \cite{KarlinerR:17} in $jj$ coupling. The spin-weighted sum of these
mass shifts in Eqs. (\ref{M2})-(\ref{M5}) vanishes: $\sum_{J}$ $Tr_{J}$($%
\Delta \mathcal{M}_{J})(2J+1)=0$, and the same holds for the eigenvalues $%
\Delta M(J,j^{\prime })$ in Eqs. (\ref{MM}). Note that the sums of
eigenvalues of $\Delta \mathcal{M}_{1/2}$ and $\Delta \mathcal{M}_{3/2}$ are
equal to the traces of the respective matrices in Eq. (\ref{MM}). Note that an alternative expression similar to the mass-shifts(\ref{MM}) is obtained recently in Ref. \cite{Ali:jh2019}.

Adding spin-independent mass $\bar{M}$, which equals to the spin-averaged
mass of the five excited $css$ systems, the baryon mass becomes $%
M(J,j_{LS}=j^{\prime })=\bar{M}+\Delta M(J,j^{\prime })$, with $\Delta
M(J,j^{\prime })$ given by Eqs. (\ref{MM}). Confronting $M(J,j_{LS}=j^{%
\prime })$ with the observed masses in Table I leads to the values of $a_{1}$%
, $a_{2}$($=\epsilon a_{1}$), $b$($=a_{1}b_{1}$) and $c$($=a_{1}c_{1}$),
with the help of the following criteria:

(i) The parameter $a_{1}$ should be positive but smaller than $119$ MeV$%
=M(\Omega _{c},3119)-M(\Omega _{c},3000)$.

(ii) The parameter $a_{2}$ is of same order with $a_{1}$ but no more than $%
a_{1}$ roughly as $a_{2}/a_{1}=\epsilon $ scales as \thinspace $m_{ss}/M_{c}$%
.

(iii) The parameter $b$ should be smaller than $a_{1}$ and $a_{2}$ as $b$
scales like $1/(m_{ss}M_{c})$. The parameter $c$ should be smallest, less
than $b$ as it scales as $P$-wave wave function near the origin.

We carry out mass analysis for all $5!=120$ a priori possible assignments of
$P$-wave $\Omega _{c}$ states and list the preferred assignments in Table I,
corresponding to the mean mass and the parameters \cite{KarlinerR:17}%
\begin{equation}
\bar{M}=3079.94 \text{MeV,}  \label{Mth}
\end{equation}%
\begin{equation}
\{a_{1},a_{2},b,c\}=\{26.96\text{, }25.76\text{, }13.51\text{, }4.04\}\text{%
(MeV),}  \label{best}
\end{equation}%
with $\epsilon =0.96,b_{1}=0.50,c_{1}=0.15$. For our preferred assignment, the inner structures of the five excited $\Omega _{c}$ baryons are

\begin{equation}
\left\{
\begin{array}{r}
\text{State\thinspace }{ |J,j}_{LS}{\small \rangle :} \\
{\small M(\Omega }_{c}{1P):} \\
\text{Main comp}.{}%
\end{array}%
\begin{array}{ccccc}
|\frac{1}{2},0^{\prime }\rangle  & |\frac{1}{2},1^{\prime }\rangle  & |\frac{%
3}{2},1^{\prime }\rangle  & |\frac{3}{2},2^{\prime }\rangle  & |\frac{5}{2}%
,2^{\prime }\rangle  \\
3000.4 & 3050.2 & 3065.6 & 3090.2 & 3119.1 \\
^{4}P_{1/2}\text{(}97\%\text{)} & ^{2}P_{1/2}\text{(}97\%\text{)} &
^{4}P_{3/2}\text{(}98\%\text{)} & ^{2}P_{3/2}\text{(}98\%\text{)} &
^{4}P_{5/2}%
\end{array}%
\right\} ,  \label{Mp}
\end{equation}%
where the third line gives the main component in terms of the normal $LS$
coupling ($^{2S+1}P_{J}$, see below). One sees that the mass degeneration
within the $J=1/2$ and $J=3/2$ multiplets removed by the flip of the charm
quark spin $\mathbf{S}_{c}$. Within the same multiplet, the lower state consists
mainly of $S=3/2$ configuration $c^{\uparrow }(s^{\uparrow }s^{\uparrow })$,
while the higher consists mainly of that with $\mathbf{S}_{c}$ flipped, $%
c^{\downarrow }(s^{\uparrow }s^{\uparrow })$ with $S=1/2$.

Expressing in terms of the $LS$ eigenstates $|^{2S+1}P_{J}\rangle $,
with $1/2\leq J\leq 5/2$ and $S=0,1$, the $Jls$ states in Eq. (\ref{Mp})
become (see Appendix B),
\begin{eqnarray}
|J &=&1/2,j^{\prime }=0^{\prime }\rangle =-0.164|^{2}P_{1/2}\rangle
+0.986|^{4}P_{1/2}\rangle ,\text{ at }3000  \notag \\
|J &=&1/2,j^{\prime }=1^{\prime }\rangle =0.986|^{2}P_{1/2}\rangle
+0.164|^{4}P_{1/2}\rangle ,\text{at }3050  \notag \\
|J &=&3/2,j^{\prime }=1^{\prime }\rangle =0.129|^{2}P_{3/2}\rangle
+0.992|^{4}P_{3/2}\rangle ,\text{at }3066  \notag \\
|J &=&3/2,j^{\prime }=2^{\prime }\rangle =-0.992|^{2}P_{3/2}\rangle
+0.129|^{4}P_{3/2}\rangle ,\text{at }3090  \notag \\
|J &=&5/2,j^{\prime }=2^{\prime }\rangle =|^{4}P_{5/2}\rangle ,\text{at }3119%
\text{. }  \label{VV}
\end{eqnarray}%
Dominately, they are  $%
|^{4}P_{1/2}\rangle ,|^{2}P_{1/2}\rangle ,|^{4}P_{3/2}\rangle
,|^{2}P_{3/2}\rangle $ and $|^{4}P_{5/2}\rangle $ with the monatomically
increasing mass, respectively, as shown in Eq. (\ref{Mp}). This is in contrast with the normal state classification of heavy baryon systems via the $jj$
coupling(see  \cite{KarlinerR:D15}, for instance),
\begin{eqnarray}
|J &=&1/2,j=0\rangle =\sqrt{\frac{1}{3}}|1^{2}P_{1/2}\rangle +\sqrt{\frac{2}{%
3}}|1^{4}P_{1/2}\rangle ,  \notag \\
|J &=&1/2,j=1\rangle =\sqrt{\frac{2}{3}}|1^{2}P_{1/2}\rangle -\sqrt{\frac{1}{%
3}}|1^{4}P_{1/2}\rangle ,  \notag \\
|J &=&3/2,j=1\rangle =\sqrt{\frac{1}{6}}|1^{2}P_{3/2}\rangle +\sqrt{\frac{5}{%
6}}|1^{4}P_{3/2}\rangle ,  \notag \\
|J &=&3/2,j=2\rangle =\sqrt{\frac{5}{6}}|1^{2}P_{3/2}\rangle -\sqrt{\frac{1}{%
6}}|1^{4}P_{3/2}\rangle ,  \label{JJS}
\end{eqnarray}%
mixing significantly between the states with $J=1/2$ and $3/2$.

It is of interest to express the mixing states (\ref{VV}) in terms of the $jj
$ bases, giving(see Appendix B)
\begin{eqnarray}
\left\vert J=\frac{1}{2},j^{\prime }=0^{\prime }\right\rangle
&=&0.711\left\vert \frac{1}{2},j=0\right\rangle +0.703\left\vert \frac{1}{2}%
,j=1\right\rangle ,  \notag \\
\left\vert J=\frac{1}{2},j^{\prime }=1^{\prime }\right\rangle
&=&-0.703\left\vert \frac{1}{2},j=0\right\rangle +0.711\left\vert \frac{1}{2}%
,j=1\right\rangle ,  \notag \\
\left\vert J=\frac{3}{2},j^{\prime }=1^{\prime }\right\rangle
&=&0.958\left\vert \frac{3}{2},j=1\right\rangle +0.286\left\vert \frac{3}{2}%
,j=2\right\rangle ,  \notag \\
\left\vert J=\frac{3}{2},j^{\prime }=2^{\prime }\right\rangle
&=&-0.286\left\vert \frac{3}{2},j=1\right\rangle +0.958\left\vert \frac{3}{2}%
,j=2\right\rangle ,  \label{Jjj}
\end{eqnarray}%
which mix almost equally between the $jj$ bases $\left\vert \frac{1}{2}%
,j=0,1\right\rangle $. Eq. (\ref{Jjj}) implies that the angular momentum $%
\mathbf{j=J-S}_{c}$ of the diquark in our mixing states with $J=1/2$ may be observed to take values either $j=0$ or $j=1$, with almost equal probabilities ($\allowbreak 0.51$ v.s. $0.49$), while those for the $J=3/2$ are mainly the $jj$ eigenstates $\left\vert
3/2,j=1,2\right\rangle $ (at $0.92$), slightly mixing the respective $\left\vert
3/2,j=2,1\right\rangle $ (at $0.08$). Hence, the $J=1/2$
states do not conserve $\mathbf{j=J-S}_{c}$ of the diquark and thereby $\mathbf{S}_{c}$. This differs the $css$ states from the
nonstrange charmed baryons, all of which can be well classified by the $jj$ eigenstates  \cite{EFG:D11,MaltmanI:D80,RobertsP:A08,KarlinerR:D15,KarlinerR:17}.

One alternative solution to assignment of the five $\Omega _{c}=css$ states
involves identification of the five masses $M(1/2,0^{\prime
}),M(1/2,1^{\prime }),M(3/2,1^{\prime }),M(3/2,2^{\prime })$, $%
M(5/2,2^{\prime })$ to be that at $3000$, $3066$, $3050$, $3090$ and $3119$ MeV, respectively. This gives rise to the parameters and the mean mass in Ref.  \cite{KarlinerR:17}
\begin{eqnarray}
\{a_{1},a_{2},b,c\} &=&\{21.40,40.75,5.67,0.44\}~\text{(MeV),}  \label{apara}
\\
\bar{M} &=&3078.23 \text{MeV.}  \label{MB2}
\end{eqnarray}%
This assignment is disfavored by unreasonable relative ratio ($1:2$) between $a_{1}$ and $a_{2}$ in the sense of the criterion (ii).

Using Eq.(\ref{MM}) further, we performed systematic search for all
remaining permutations among all $5!$ possibilities and find no acceptable
solution for considered permutations: the parameters obtained either have
negative (unacceptable) signs of $a_{1}$ and $a_{2}$ or unreasonable values
in the light of the criteria (i)-(iii).

\begin{table}[htb]
\caption{Mass and parameters for two $J^{P}$ assignments (\ref{Mp}) and the alternative arrangement of four LHCb-observed masses of the excited $\Omega _{c}$ systems.  All $2\ast 5=10$ possible choices of mass inputs out of the five measured
masses are considered. The predicted mass at presumably unseen state is marked in square bracket. All parameters except the Regge slope $a$ in MeV.  \label{Table 2}}
\[\begin{array}{cccccccccccc}
\hline
\hline

|\frac{1}{2}, 0^{\prime }\rangle   &|\frac{1}{2}, 1^{\prime }\rangle    &|\frac{3}{2}, 1^{\prime }\rangle& |\frac{3}{2}, 2^{\prime }\rangle   &|\frac{5}{2}, 2^{\prime }\rangle    &a_{1}
&a_{2}       &b          &c        &a( GeV^2)       &\bar{M}(1P)      &\bar{M}(2S) \\\hline
$[2995.0]$ &$3050$ &$3066$  &$3090$     &$3119$     &
$27.5$ & $27.0$   & $15.5$ & $3.6$    &\multicolumn{1}{c}{$0.316$} & $3079$ & $3244$ \\

$3000$ &$[3049.0]$     &$3066$  &$3090$     &$3119$ &
$27.2$ &$25.2$     &$13.7$  & $4.4$   &\multicolumn{1}{c}{$0.316$} & $3080$ & $3244$ \\

$3000$ &$3050$     &$[3068.2]$  &$3090$ &$3119$     &
$26.7$ & $24.8$    &$15.4$ & $5.0$    &\multicolumn{1}{c}{$0.317$} & $3081$ & $3245$ \\

$3000$ &$3050$     &$3066$  &$[3095.4]$ &$3119$      &
$28.2$ & $23.1$    &$14.4$  &$2.3$    &\multicolumn{1}{c}{$0.317$} & $3081$ & $3246$ \\

$3000$ &$3050$     &$3066$  &$3090$     &$[3115.6]$     &
$26.3$ & $23.7$    &$14.7$  &$3.2$   &\multicolumn{1}{c}{$0.315$} & $3079$ & $3243$ \\

$\lbrack 3000.4]$ &$3066$&$3050$&$3090$ &$3119$     &
$21.4$ & $40.8$ & $5.7$ & $0.44$ &\multicolumn{1}{c}{$0.314$} & $3078$ & $3242$ \\

$3000$ &$[3067.4]$      &$3050$ &$3090$   &$3119$   &
$20.4$ & $41.9$ & $6.4$ & $1.2$ &\multicolumn{1}{c}{$0.315$} & $3078$ & $3242$ \\

$3000$ &$3066$     &$[3051.0]$  &$3090$ &$3119$     &
$21.4$ & $40.4$    &$6.1$ & $0.52$    &\multicolumn{1}{c}{$0.315$} & $3078$ & $3242$ \\

$3000$ &$3066$     &$3050$  &$[3090.1]$ &$3119$      &
$21.3$ & $40.8$    &$5.7$  &$0.59$    &\multicolumn{1}{c}{$0.314$} & $3078$ & $3242$ \\

$3000$ &$3066$     &$3050$  &$3090$     &$[3117.5]$     &
$21.4$ & $39.7$    &$5.7$  &$-0.57$   &\multicolumn{1}{c}{$0.314$} & $3078$ & $3241$ \\

\hline\hline
\end{array}\]
\end{table}

We also made a "recovery" test by fitting four out of the five observed $\Omega
_{c}$ masses for the $J^{P}$ assignments (\ref{Mp}) and the alternative solution with $M(1/2,1^{\prime})$ and $M(3/2,1^{\prime})$ interchanged, corresponding to the parameters (\ref{apara}), to see
if the remaining measured mass, unused in fitting inputs, can be reproduced at the site of given spin-states(marked by square bracket). The results for the "recovery" test are shown in Table II for all $2\ast 5=10$ possibilities of state arrangements for chosen inputs and indicate that the test successes for both, whereas the parameter values clearly favor the assignment (\ref{Mp}) in the light of the criterion (ii): $a_{2}$ is notably large over $a_{1}$, nearly two times,  for the alternative solution.

One can also use the mass scaling of the spin-interaction couplings, from $D_{s}$ mesons to the $css$ baryons, to explain the values (\ref{best}) for our preferred assignment. Using the scaling relation \cite{KarlinerR:D15,JiaLH:19}, one finds
\begin{eqnarray}
a_{1}(css) &=&a_{1}(c\bar{s})\left( \frac{m_{s}}{m_{ss}}\right) =(89.4~ \text{%
MeV})\left( \frac{328}{991}\right) =29.6~ \text{MeV,}  \label{scal1} \\
a_{2}(css) &=&\frac{a_{2}(c\bar{s})}{1+m_{ss}/M_{c}}=\frac{40.7~ \text{MeV}}{%
1+991/1440}=24.1 ~\text{MeV,}  \label{scal2}
\end{eqnarray}%
which are close to the values (\ref{best}), but away from that in (\ref%
{apara}). Here, $a_{1,2}(c\bar{s})$ are the parameters of the spin-orbit
interactions for the P-wave $D_{s}=c\bar{s}$ mesons, with the respective
values $89.4$ MeV and $40.7$ MeV \cite{KarlinerR:D15}. The masses($m_{s}=328$%
MeV and $M_{c}=1440$ MeV) of the strange quark $s$ and  the charm quark are
from Regge trajectory fit of the $D_{s}$ mesons and the charmed baryons $%
\Sigma _{c}/\Xi _{c}^{\prime }$ \cite{JiaLH:19} while the mass($m_{ss}=991$%
 MeV) of the ss-diquark comes from the Regge trajectory fit of the five
excited $\Omega _{c}$ states in Table I(cf. the discussion following (\ref%
{RT}) in Sect. III).

\section{Alternative possibilities examined}
The possibility that must be checked is that not all of the $\Omega
_{c}$ states in Table I are relative $P$-wave excitations between $ss$
diquark and the charmed quark  \cite%
{EFG:D11,ChenHL:D17,WangXZZ:D17,ChengC:D17,AgaevAS:D17}. For instance, two
$2S$ wave candidates are predicted for the $c(ss)$ at masses $3088$ MeV ($J^{P}=1/2^{+}$) and $3123$ MeV($J^{P}=3/2^{+}$) in Ref. \cite{EFG:D11}, not far from two higher masses($3090$ and $3119$ MeV) reported by LHCb. This leaves
possibility that some of the lower states may be that out of the five states
of $P$-waves, with other $P$-wave states unseen somehow, probably due to
locating below the $\Xi _{c}^{+}K^{-}$ threshold ($\simeq 2962$ MeV), or
invisible due to the near degeneracies \cite{EFG:D11}. To disentangle this
possibility, we extend the mass analysis in Sect. II by assuming one of the
five measured masses of the $\Omega _{c}$ states in Table I to be that of 2S
state and comparing it with mass estimation of the 2S state with the help
of Regge trajectory \cite{JiaDH:Q18,JiaD:E19}. The results suggest that all
measured masses of the excited $\Omega _{c}$ states are too low to be a 2S
state.

We use the mass shift formula (\ref{MM}) to perform mass analysis for all $5!$
permutations of the $J^{P}$ arrangement with one presumed 2S state removed from
five mass inputs. Apart from the $10$ arrangements already
given in Table II, we list, in Table III, the other candidate solutions with the
parameters that are not too far away from the criteria (i)-(iii), and the thereby predicted P-wave mass(presumably unseen) enclosed in square bracket. Though two arrangements in the first and second data rows recover approximately the observed masses absent in the inputs, all of these permutated arrangements are disfavored by the criteria (i)-(iii), having unreasonable values of the parameters listed: either $a_{2}$ or $b$ is notably large over $a_{1}$ or negative, or $c$ is abnormally large compared with one of $a_{2}$ and $b$, or both.

\begin{table}[htb]
\caption{Parameters and spin-averaged masses (MeV) for the selected
permutations of $J^{P}$ assignments of the four states out of excited $%
\Omega _{c}$ levels. Only the arrangements that leads to the parameters near
to reasonable values are listed among $5\ast 4!=120$ possible permutations
of mass inputs considered, with the mass prediction at unseen level marked by
square bracket. The shown also includes the ensuing prediction for the
Regge slope $a$ (see text, in GeV$^{2}$) and spin-averaged 2S-wave mass(in MeV) of the $c(ss)$ system. $m_{ss}=991$ MeV.  \label{Table 3}}
\[\begin{array}{cccccccc}
\hline\hline

|\frac{1}{2},0^{\prime }\rangle,|\frac{1}{2},1^{\prime }\rangle, |\frac{3}{2},1^{\prime }\rangle, |\frac{3}{2},2^{\prime }\rangle, |\frac{5}{2},2^{\prime }\rangle   & a_{1} & a_{2} &b &c &a($ GeV$^{2}) & \bar{M}(1P) & \bar{M}(2S) \\ \hline
3000,[3052.3],3066,3119,3090 & 23.0 & -2.19 & 35.0 & -3.43 &0.313&3077.0 & 3240.3\\
3000,3066,3050,3119,[3093.9] & 27.8 & 7.86 & 11.0 & -23.9 &0.312&3076.2 &3239.3 \\
3050,3066,3000,[3082.2],3119 & 10.7 & 49.9 & -36.2 & 10.6&0.307&3070.9 & 3231.7\\
3000,3090,3050,[3102.5],3119 & 14.6 & 48.9 & 10.1 & -3.20 &0.320&3083.7&3249.9\\
3066,[3108.6],3000,3119,3090 & 11.2 & 24.8 & -49.1 & -38.41&0.312 &3076.0&3238.9\\
$[3008.0]$,3000,3050,3066,3119 & 32.3 & 14.2 &5.77 & 28.9 &0.302 &3066.4 & 3225.2 \\
3050,3066,3000,[3078.7],3119 & 11.1 & 47.6 & -38.8 &12.7&0.306& 3074.1 &3230.6\\
\hline\hline
\end{array}\]
\end{table}

\medskip

The mass estimation of the $\Omega _{c}$ systems in $2S$ wave can be given
utilizing a Regge-like relation for the spin-averaged mass $\bar{M}$ of the charmed baryons \cite{BChen:A15,JiaLH:19}%
\begin{equation}
(\bar{M}-M_{c})^{2}=\pi aL+\left[ m_{d}+M_{c}\left( 1-\frac{m_{barec}^{2}}{%
M_{c}^{2}}\right) \right] ^{2},  \label{RT}
\end{equation}%
in which $a$ is the Regge slope, $m_{d}$ is the light diquark mass involved,
$m_{barec}=1.275$ GeV the bare mass of charm quark and $L$ the orbital
angular momentum of the systems. The charm quark mass $M_{c}=1.44$ GeV is
determined by confronting the relation (\ref{RT}) with the charmed baryons $%
\Sigma _{c}/\Xi _{c}^{\prime }$ \cite{JiaLH:19}. In Ref. \cite%
{JiaD:E19}, a trajectory slope ratio $\pi :2$ for the
radially and angular excitations is suggested in applying Eq. (\ref{RT}) to the
heavy mesons $B/D$'s and $B_{s}/D_{s}$'s.

\begin{figure}[!h]
\centering
\begin{tabular}{c}
\includegraphics[width=0.44\textwidth]{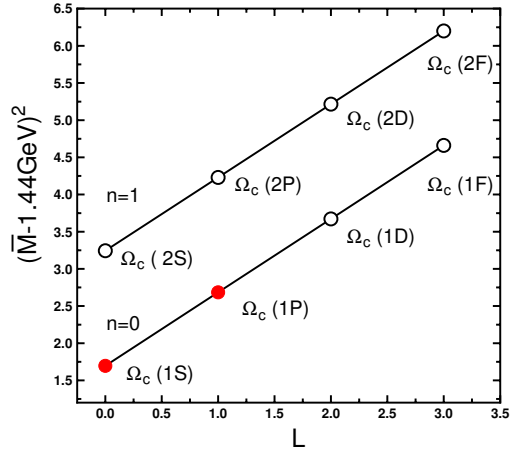}
\end{tabular}
\caption{(Color online). Regge trajectories of the $\Omega_c$ baryons relating the shifted spin-averaged mass squared to the orbital angular momentum $L$ of the systems, with the parameters $m_{d=ss}=991$ MeV and the Regge slope $a=0.316$GeV$^{2}$ corresponding to our assignment (\ref{Mp}). The (red) solid circles correspond to the observed (mean) masses. } \label{figRegge}

\end{figure}
\medskip

By analogy of the $c(ss)$ system with $D_{s}\,=c\bar{s}\,$\ mesons, we
estimate the 2S-wave mass of the $c(ss)$ state using Eq. (\ref{RT}) with $a$
replaced by $a(\pi /2)$ and $L$ by $n$, the radial quantum number of the $%
c(ss)$ system,
\begin{equation}
\bar{M}(2S)=M_{c}+\sqrt{\pi a(\pi /2)+\left( m_{d}+M_{c}\left( 1-\frac{%
1.275^{2}}{M_{c}^{2}}\right) \right) ^{2}}\text{,}  \label{M2S}
\end{equation}%
where $m_{d}=0.991$ GeV and the Regge slope $a$ (listed in Table II and
III) are solved from (\ref{RT}) applying to the spin-averaged masses $\bar{M}%
(1S)$ and $\bar{M}(1P)$ for a given assignment, and $M_{c}=1.44$ GeV. Here,
the observed lowest masses of the $\Omega _{c}$, $M(\Omega
_{c},1/2^{+})=2695.2\pm 1.7$ MeV and $M(\Omega _{c},3/2^{+})=2765.9\pm 2.0$ MeV%
 \cite{Tanabashi:D18}, lead to their spin-averaged mass in S-wave
\begin{equation}
\bar{M}(\Omega _{c},1S)=\frac{1}{6}\left( 2M(1/2^{+})+4M(3/2^{+})\right)
=2742.3\pm 1.9~\text{MeV.}  \label{Omv}
\end{equation}%
Usage of the mean mass (\ref{Mth}) in P-wave($L=1$) and the mass (\ref{Omv}) in S-wave($L=0$) in Eq. (\ref{RT}) gives
\begin{equation}
\Omega _{c}:m_{d}=m_{ss}=0.991\text{ GeV, }a=0.316\text{ GeV}^{2}\text{.}
\label{aMN}
\end{equation}

Given trajectory parameters, one can predict Regge trajectories of the excited baryons $\Omega _{c}$(FIG. 1) with radial quantum number $n=0$ and $1$. In Table IV in which $n$ stands for $n+1=1,2$, we list corresponding spin-averaged masses of the excited $\Omega _{c}$'s predicted by Eqs. (\ref{RT})and (\ref{M2S}), which are quite helpful for further constructing whole families of the baryon $\Omega _{c}$'s.

\smallskip

\begin{table}[htb]
\caption{Mean masses of the excited $\Omega _{Q}$'s predicted by Eqs. (\ref{RT}) and (\ref{M2S}).   \label{Table 4}}
\[\begin{array}{cccccc}
\hline\hline
State(\text{MeV}) &  & \bar{M}(nS) & \bar{M}(nP) & \bar{M}(nD) & \bar{M}(nF) \\ \hline
\Omega _{c}($n=1$) &  & 2742.3 & 3078.3 & 3356.2 & 3598.7 \\
\Omega _{c}($n=2$) &  & 3241.1 & 3496.5 & 3723.6 & 3930.0 \\
\Omega _{b}($n=1$) &  & 6051.0 & 6342.0 & 6593.3 & 6817.7 \\
\Omega _{b}($n=2$) &  & 6489.2 & 6724.1 & 6936.6 & 7132.4 \\
\hline\hline
\end{array}\]
\end{table}

Depending on state arrangements having different $P$-wave mean-mass, the above values vary slightly. The 2S-wave masses of the $c(ss)$ system estimated via (\ref{M2S}) are listed in Table II and III for given
assignments. Evidently, the 2S-wave mass $\bar{M}(2S)\geq 3220$ MeV, and
both of the higher mass states at $3090$ MeV and $3119$ MeV are too low to be a $2S$ state candidates. The same conclusion holds also for other low-lying masses, as shown in Table II and III.

For other assignments searched, no solutions of the predicted unseen
states, below the $2962$ MeV threshold (of $\Xi _{c}^{+}K^{-}$) or near
degeneracies with other masses, are found up to requirements (i)-(iii) in
Sect. II.

\section{P-wave masses for $\Omega _{b}=b(ss)$ Systems}
Recent observation of four $\Omega _{b}^{-}$ states \cite{Aaij:L2020}
around $6.3$ GeV(Table I) makes it timely to apply mass analysis of the $%
\Omega _{c}=c(ss)$ states to the bottom systems $b(ss)$ which consists of a
bottom quark $b$ and a spin-1 diquark $ss$. Heaviness of the $b$ quark
implies that Eqs. (\ref{MM}) should certainly be applicable for the P-wave $\Omega
_{b}^{-}$ states. We use it to search the possible assignments with
the following inputs:

(i) The parameter $a_{1}$ is positive and no more than the maximum gap
between the observed masses : $a_{1}<2\ast (6349.88-6315.64)$ $=68.48$ MeV.
(ii) The parameter $a_{2}$ is of same order of $a_{1}$ roughly in magnitude,
but should be smaller than $a_{1}$.
(iii) The tensor parameter $b$ should be smaller than $a_{1}$ in magnitude.
(iv) The hyperfine parameter $c$ is set to zero.

Assuming one of P-wave states is unseen experimentally the search is performed systematically via matching the observed masses of the $\Omega _{b}$ states in Table I with that in Eq. (\ref{MM}) for all $4!=24$ a priori possible $J^{P}$ assignments of P-wave states and $5$ possible permutations of the unseen $P$-wave state that is expected to exist somewhere around $6.3$ GeV.

The most preferred results are achieved by the identification of four
masses at $6315.64,6330.30$, $6339.71$ and $6349.88$ MeV as $M(1/2,0^{\prime
})$, $M(1/2,1^{\prime })$, $M(3/2,1^{\prime })$, $M(3/2,2^{\prime })$. The
corresponding parameters and the mean mass are
\begin{equation}
a_{1}=8.98~\text{MeV, }a_{2}=4.11~\text{MeV, }b=7.61~\text{MeV, }\bar{M}=6342.0%~
~\text{MeV,}  \label{a11}
\end{equation}%
which lead, by Eqs. (\ref{MM}), to the following $J^{P}$ assignment,
\begin{equation}
\left\{
\begin{array}{c}
\text{State(}J^{P}\text{):} \\
M(\Omega _{b},1P\text{):}%
\end{array}%
\begin{array}{ccccc}
|1/2,0^{\prime }\rangle  & |1/2,1^{\prime }\rangle  & |3/2,1^{\prime
}\rangle  & |3/2,2^{\prime }\rangle  & |5/2,2^{\prime }\rangle  \\
\multicolumn{1}{r}{6315.4,} & \multicolumn{1}{r}{6332.0} &
\multicolumn{1}{r}{6337.8} & \multicolumn{1}{r}{6350.0} & \multicolumn{1}{r}{
6351.5^{\text{Pd}}}%
\end{array}%
\right\} .  \label{Momi}
\end{equation}%
In terms of the $LS$ basis, their inner structures are
\begin{eqnarray}
|1/2,0^{\prime }\rangle  &=&-0.190|^{2}P_{1/2}\rangle
+0.982|^{4}P_{1/2}\rangle ,\text{at }6316,  \notag \\
|1/2,1^{\prime }\rangle  &=&0.982|^{2}P_{1/2}\rangle
+0.190|^{4}P_{1/2}\rangle ,\text{ at }6330,  \notag \\
|3/2,1^{\prime }\rangle  &=&0.488|^{2}P_{3/2}\rangle
-0.873|^{4}P_{3/2}\rangle ,\text{at }6340,  \notag \\
|3/2,2^{\prime }\rangle  &=&0.873|^{2}P_{3/2}\rangle
+0.488|^{4}P_{3/2}\rangle ,\text{at }6350,  \notag \\
|5/2,2^{\prime }\rangle  &=&|^{4}P_{5/2}\rangle \text{ at }6351.5^{\text{Pd}}%
\text{. }  \label{In0}
\end{eqnarray}%
which are mainly the $P$-wave states $|^{4}P_{1/2}\rangle
,|^{2}P_{1/2}\rangle $, two mixings of $|^{4}P_{3/2}\rangle \ $and $%
|^{2}P_{3/2}\rangle $, and $|^{4}P_{5/2}\rangle $, respectively.

The experimental missing of the $J=5/2$ state in Table I at $M(5/2,2^{\prime })=6351.5$, which we predict, is most likely due to its degeneracy with the nearby state at $M(3/2,2^{\prime })=6350$ MeV. If this is the
case, it may be hidden in the observed peak around $6350$ MeV, which, though
appears consistent with a single resonance, is actually composed of two.
This is compared to the recent assignment in Ref. \cite%
{KarlinerR:Ax20}, where the mass for the unseen $J=5/2$ state ranges from $6355$ MeV to $6383$ MeV or from $6380$ MeV to $6407$ MeV, with the favored parameters $a_{1}=10.20$ MeV, $a_{2}+c=10.04$ MeV and $b=4.75$ MeV.

One of other plausible solutions consists of identification of the four
masses at $6315.64,6330.30$, $6339.71$ and $6349.88$ MeV as $|1/2,0^{\prime
}\rangle ,|3/2,1^{\prime }\rangle $,$|3/2,2^{\prime }\rangle $ and $%
|5/2,2^{\prime }\rangle $. This assignment gives parameters and
spin-averaged mass(in MeV)
\begin{equation}
\{a_{1},a_{2},b\}=\{7.39\text{, }10.32\text{, }1.93\}\text{, }\bar{M}=6337.4%
\text{,}  \label{aab}
\end{equation}%
where the parameter $a_{2}=10.32$ MeV is larger than $a_{1}=7.39$ MeV, not
compatible with the criterion (ii).

There are still two additional possibilities involving identification of the
four observed $\Omega _{b}$ states at higher $j^{\prime }=1^{\prime
},2^{\prime }$. The first is to identify the states at $6330$, $6316,6340$
and $6350$ MeV with the respective states at $M(1/2,1^{\prime })$, $%
M(3/2,1^{\prime })$, $M(3/2,2^{\prime }),M(5/2,2^{\prime })$, giving the
parameter set(in MeV)
\begin{equation}
\{a_{1},a_{2},b\}=\{10.86\text{, }6.70\text{,}-10.54\}\text{, }\bar{M}=6333.6%
\text{,}  \label{Pad1}
\end{equation}%
and the presumably unseen state at $M(1/2,0^{\prime })=6311.5$ MeV. This lowest state, if exist, are most likely to be visible. Together with abnormal (by the criteria (iii)) value $|b|\simeq 11$ MeV compared to $a_{1}=10.86$ MeV, we disfavor this assignment.

The second is to interchange two states at $6340$ and $6350$ MeV in the
assignment above, leading to the parameter set(in MeV)
\begin{equation}
\{a_{1},a_{2},b\}=\{11.93\text{, }-8.98\text{,}-2.92\}\text{, }\bar{M}=6331.7%
\text{,}  \label{Pad2}
\end{equation}%
and the five states $|1/2,j^{\prime }=0^{\prime }\rangle $ at $6304.7$%
(prediction), $|1/2,j^{\prime }=1^{\prime }\rangle $ at $6330.3$, $%
|3/2,j^{\prime }=1^{\prime }\rangle $ at $6315.6$, $|3/2,j^{\prime
}=2^{\prime }\rangle $ at $6349.9$ and $|5/2,2^{\prime }\rangle
=|^{4}P_{5/2}\rangle $ at $6339.7$ MeV. Away(about $10$ MeV) from the observed
masses $6315$ MeV, the presumed missing state, at $M(1/2,0^{\prime })=6304.7$ MeV, should not be hidden in any of the four observed peaks by LHCb. The unacceptable(negative) value of $a_{2}=-8.98$ MeV also disfavors this assignment.

\begin{table}[htb]
\caption{Parameters and masses(all in MeV) for the excited $\Omega _{b}$ when one of four observed masses was removed from the mass inputs. The removed one (enclosed by square bracket) is reproduced by Eq. (\ref{MM}) approximately.   \label{Table 5}}
\[\begin{array}{ccccc}
\hline\hline
|\frac{1}{2},0^{\prime }\rangle ,\ |\frac{1}{2},1^{\prime }\rangle ,\ |%
\frac{3}{2},1^{\prime }\rangle ,\ |\frac{3}{2},2^{\prime }\rangle ,|\frac{5}{%
2},2^{\prime }\rangle  & a_{1} & a_{2} & b & \bar{M}(1P) \\ \hline
[6314.5],6332.0,6337.8,6350.0,6351.5 & 9.02 & 4.44 & 7.92 & 6341.8 \\
6315.4,[6332.1],6337.8,6350.0,6351.5 & 8.91 & 4.27 & 7.53 & 6342.0 \\
6315.4,6332.0,[6337.7],6350.0,6351.5 & 8.95 & 4.25 & 7.48 & 6341.9 \\
6315.4,6332.0,6337.8,[6350.5],6351.5 & 8.99 & 4.01 & 7.12 & 6342.1\\

\hline\hline
\end{array}\]
\end{table}

\begin{table}[htb]
\centering%
\caption{Parameters and spin-averaged masses(in MeV) for the selected
permutations of $J^{P}$ assignments of the four excited $\Omega _{b}$
levels, with predicted mass at unseen level(marked by square bracket). All
arrangements leading to highly unreasonable or unacceptable parameters are
not listed. The shown includes also ensuing prediction for the Regge slope $a
$ (see text) and spin-averaged $css$ masses in $2S$ wave. $m_{ss}=991$ MeV.
We denoted the five states $|\frac{1}{2},0^{\prime }\rangle ,|\frac{1}{2}%
,1^{\prime }\rangle ,|\frac{3}{2},1^{\prime }\rangle ,|\frac{3}{2},2^{\prime
}\rangle ,|\frac{5}{2},2^{\prime }\rangle $ by $|1\rangle ,|2\rangle
,|3\rangle ,|4\rangle ,|5\rangle $ for short, respectively. \label{Table 6}}
\[\begin{array}{cccccccccccc}
%\hline
%\multicolumn{5}{c}{\text{State arrangement}}\\%
\hline\hline
|1\rangle &|2\rangle &|3\rangle &|4\rangle&|5\rangle&a_{1}&a_{2} &b &a($ GeV$^{2})&\bar{M}(1P)&\bar{M}(2S)\\\hline
$[6312.4]$&6316&6329&6340&6350& 21.2& $26.9$ &42.4&\multicolumn{1}{c}{$0.297$} & $6324.2$ & $6463.3$ \\
6314&[6331.4]&6330&6341&6350&7.39& $10.3$ & $1.93$ &\multicolumn{1}{c}{$0.313$} & $6337.4$ & $6482.6$ \\
6316&6330&[6321.8]&6340&6350& $7.17$ & $13.3$ & $-3.28$ &\multicolumn{1}{c}{$0.310$} & $6335.4$ & $6479.7$ \\
6316&6331&6338&[6388.1]&6350 & $12.0$ & $-13.9$ & $24.1$ &\multicolumn{1}{c}{$0.328$} & $6350.1$ & $6501.2$ \\
6313&6332&6338&6351&[6352.5] & $9.4$ & $4.87$ & $8.37$ &\multicolumn{1}{c}{$0.318$} & $6342.3$ & $6489.8$ \\
$[6311.5]$&6330&6316&6340&6350 & $10.9$ & $6.70$&$-10.5$ &\multicolumn{1}{c}{$0.308$} &6333.6& $6477.1$\\
$[6304.7]$&6330&6316&6350&6340& $11.9$ & $-8.98$ &-2.92&\multicolumn{1}{c}{$0.306$} & $3331.7$ & $6474.4$ \\
\hline\hline
\end{array}\]
\end{table}

In addition, we carry out "recovery" test (Table V, all in MeV) for the excited $\Omega _{b}$'s, in which one of four observed masses is reproduced approximately by Eq. (\ref{MM}) at the site enclosed by square bracket when it was removed from four mass inputs. The uncertainty is about $2$  at most. This further confirms our prediction $M(5/2,2^{\prime})=6351.5$ MeV for the unseen (excited) $\Omega _{b}$ state.

Could it be possible that some higher states of these $\Omega _{b}$ peaks
are the $2S$ excitations? Our answer to this question is negative. We employ Eq. (\ref{MM}) to perform mass fitting for the four mass inputs of the $\Omega _{b}$ baryons in Table I to find the three parameters $a_{1,2}$, $b$ and $\bar{M}(1P)$, and thereby estimate the 2S-wave mass of the $\Omega _{b}$ baryons, shown in Table VI, using Eq. (\ref{M2S}) with $M_{c}$ replaced by $M_{b}=4.48$ GeV (the charm bare mass $1.275$ GeV by that of the bottom quark $4.18$ GeV). For all assignments with the parameters not far away from the inputs (i)-(iv), we find
\begin{equation}
\bar{M}(\Omega _{b},2S)\gtrsim 6450~\text{MeV,}  \label{Ob2S}
\end{equation}
which disfavors identifying newly LHCb-reported $\Omega _{b}$ to be a $2S$ excitation.  During calculating, the $ss$ diquark mass, $m_{d=ss}=1.001$ GeV, is determined from Eq. (\ref{RT}) applied to spin-averaged
mass $M(\Omega_{b},1S)=6061.4$ MeV of the two lowest $1S$ states, the $\Omega _{b}(1/2^{+})$ at $6046.1\pm 1.7$ MeV and its partner $\Omega _{b}^{\ast }$($3/2^{+}$) with mass $6046.1+$ $\Delta E(\Omega _{b},1S)$. Here, $\Delta
E(\Omega _{b},1S)$ stands for the $1S$ level-splitting between the $\Omega_{b}^{\ast }$($3/2^{+}$) and $\Omega _{b}(1/2^{+})$, and can be estimated from the corresponding $1S$ level-splitting of the ground states $\Omega _{c}$ via the scaling \cite{KarlinerR:D15},
\begin{eqnarray*}
\Delta E(\Omega _{b}) &=&\left( \frac{M_{c}}{M_{b}}\right) \Delta E(\Omega
_{c}) \\
&=&\left( \frac{1.44}{4.48}\right) (71~\text{MeV)} \\
&\simeq &23~\text{MeV.}
\end{eqnarray*}%
which yields the spin-averaged mass in $1S$-wave:
\begin{eqnarray*}
\bar{M}(\Omega _{b},1S) &=&\frac{1}{6}[6046.1\times 2+(6046.1+23)\times 4] \\
&=&6061.4~\text{MeV.}
\end{eqnarray*}

Hence, Eq. (\ref{Ob2S}) disfavors the $2S$ state assignment for any of four observed $\Omega_{b}$ states, as the 2S state is too high in levels to accommodate all $\Omega _{b}$ states in Table I. Similar analysis giving Eq. (\ref{aMN}) applies to the P-wave bottom baryons $\Omega _{b}$ with the mean mass in Eq. (\ref{a11}) and gives $a=0.318$ GeV$^{2}$. This leads to the mean masses in Table IV for the $\Omega _{b}$'s, by which one can plot Regge trajectories (FIG. 2) for them.

\begin{figure}[!h]
\centering
\begin{tabular}{c}
\includegraphics[width=0.44\textwidth]{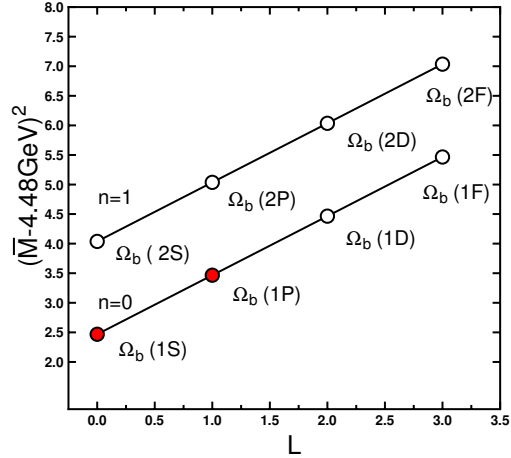}
\end{tabular}
\caption{(Color online). Regge trajectories of the $\Omega_b$ baryons relating the shifted spin-averaged mass squared to the orbital angular momentum $L$ of the systems, with the parameters $m_{d=ss}=991$ MeV and the Regge slope $a=0.318   GeV^{2}$ corresponding to our assignment(\ref{Mp}). The (red) solid circles correspond to the observed (mean) masses. } \label{figRegge2}

\end{figure}

\section{Excited Baryons $\Sigma_{c}/\Xi_{c}^{\prime}$ and $\Sigma_{b}/\Xi_{b}^{\prime}$ : P-wave Masses}
It is possible to apply the mixing $Jls$ coupling to computation of the P-wave masses of the less strange heavy baryons($\Sigma _{Q}/\Xi _{Q}^{\prime }$), for which only a single measured mass available for each one of the odd-parity baryons, $2801$$/2942.3$ MeV for the $\Sigma _{c}(2800)/\Xi_{c}^{\prime }(2942)$ and $6098.0$$/6226.9$ MeV for the $\Sigma _{b}(6097)$/$\Xi _{b}^{\prime }(6227)$ (Table I, or see \cite{Tanabashi:D18}). The basic idea for such a computation is to choose our previous estimates of the parameters $a_{1,2}$, $\bar{M}$ for the $1P$ states\cite{JiaLH:19}, as well as the following rough estimate of the parameter $c$(with $m_{nn}=745$ MeV, $m_{ns}=872$ MeV)
\begin{equation}
\text{Initial: }c(cqq)\approx c(css)\left( \frac{m_{ss}}{m_{qq}}\right)
=\left\{
\begin{array}{rr}
5.37 \text{ MeV}, & \Sigma _{c}, \\
4.59 \text{ MeV}, & \Xi _{c}^{\prime }.%
\end{array}%
\right.   \label{cqq}
\end{equation}%
with $c(css)=4.04$ MeV given in Eq. (\ref{best}), to be three initial inputs to give a $b$-dependent P-wave mass for each baryon using the nonlinear relation (\ref{MM}), and then find optimal fit of $b$ as the initial inputs via matching $b$-dependent mass with the measured one for each baryons(Table I). Finally, one
evolves the obtained initial inputs via the relation (\ref{MM}) to find stable
values of the five parameters. We list all initial parameters including $%
\bar{M}(1P)$ in the Table VII, where
\begin{table}[htb]
\caption{Initial parameters and mean masses(all in MeV) for the less strange heavy baryons($\Sigma _{c,b}/\Xi _{c,b}^{\prime }$).   \label{Table 7}}
\[\begin{array}{cccccc}
\hline\hline
\text{Initial input}  & a_{1} & a_{2} & b & c & \bar{M}(1P) \\ \hline
\Sigma _{c}(\text{MeV})  & 39.4 & 26.8 & 20.1 & 5.37 & 2774.1 \\
\Xi _{c}^{\prime }(\text{MeV})  & 33.6 & 25.3 & 17.9 & 4.59 & 2923.0 \\
\Sigma _{b}(\text{MeV})  & 12.7 & 8.61 & 6.45 & 1.73 & 6088.4 \\
\Xi _{b}^{\prime }(\text{MeV})  & 10.8 & 8.10 & 5.76 & 1.48 & 6248.2 \\
\hline\hline
\end{array}\]
\end{table}
the two inputs $a_{1}=39.4$ MeV/$33.6$ MeV for the P-wave baryons $%
\Sigma _{c}/\Xi _{c}^{\prime }$ are scaled from that $a_{2}=26.8$ MeV/$25.3$ MeV for the $D_{s}(1P)$ masses in Ref. \cite{JiaLH:19}, and the mean-masses $\bar{M}=2774.1$ MeV/$2923.0$ MeV for the P-wave baryons $\Sigma _{c}/\Xi _{c}^{\prime }$ are taken from Table III of Ref. \cite{JiaLH:19}, which are extrapolated from Regge-trajectory of the $\Lambda _{c}/\Xi _{c}$ spectra.

For the P-wave bottom baryons $\Sigma _{b}$/$\Xi _{b}^{\prime }$, corresponding initial inputs (Table VII) follow from the following estimations: the initial parameter $c$ is estimated, by scaling from the $\Omega _{c}$ to the $\Omega _{b}$ and then to the bottom baryon $bqq$, to be about
\begin{equation}
\text{Initial: }c(bqq)\approx c(css)\left( \frac{M_{c}}{M_{b}}\right) \left(
\frac{m_{ss}}{m_{qq}}\right) =\left\{
\begin{array}{rr}
1.73\text{ MeV,} & \Sigma _{b}, \\
1.48\text{ MeV,} & \Xi _{b}^{\prime }.%
\end{array}%
\right.   \label{bqq}
\end{equation}%
with $M_{c}=1.44$ GeV and $M_{b}=4.48$ GeV[37]. The mass scaling $%
a_{2}(bss)\simeq a_{2}(css)(M_{c}/M_{b})$ is used to estimate $a_{2}$ while the HQS relation $a_{1}(bqq)=a_{1}(cqq)$ in Ref. [35] needs to be
corrected, as indicated by $a_{1}(bss)=8.98<26.96=a_{1}(css)$ and also by failure
to find reasonable parameters if setting $a_{1}(\Sigma _{b})= 39.4$ MeV$%
=a_{1}(\Sigma _{c})$ or $a_{1}(\Xi _{b}^{\prime })= 33.6$ MeV$=a_{1}(\Xi
_{c}^{\prime })$. We find that choosing initial values of $a_{1}(bqq)$ to be
about one third of $a_{1}(cqq)$, as $a_{1}(bss)\simeq a_{1}(css)/3$
indicated, yields reasonable and stable values of the parameters during
iteration of Eq. (\ref{MM}). In the Table VIII($\Diamond $ denotes mass observed), we show the results of iteration starting with the initial inputs in Table VII,
with reasonable values of the parameters achieved. Note that the obtained
values of $b$ is about one half of that of $a_{1}$.

\begin{table}[htb]
\caption{Masses and parameters(all in MeV) refined from the initial inputs in Table VII via iteration of the nonlinear relation (\ref{MM}) for the less strange heavy baryons $\Sigma _{c}/\Xi _{c}^{\prime }$ and $\Sigma _{b}/\Xi _{b}^{\prime }$ in P-wave.   \label{Table 8}}
\[\begin{array}{ccccccc}
\hline\hline
State: & |\frac{1}{2},0^{\prime }\rangle ,\ |\frac{1}{2},1^{\prime }\rangle
,\ |\frac{3}{2},1^{\prime }\rangle ,\ |\frac{3}{2},2^{\prime }\rangle ,|%
\frac{5}{2},2^{\prime }\rangle  & a_{1} & a_{2} & b & c & \bar{M}%
(1P) \\   \hline
\Sigma _{c}(\text{MeV}) & 2668.4,2723.1,2757.3,2801.0^{\Diamond },2826.6 & %
39.96 & 21.75 & 20.70 & 7.85 & 2776.4 \\
\Xi _{c}^{\prime }(\text{MeV}) & 2840.6,2881.6,2908.9,2942.3^{\Diamond },2969.5
& 32.89 & 20.16 & 16.50 & 7.17 & 2925.9 \\
\Sigma _{b}(\text{MeV}) & 6053.9,6071.8,6082.8,6098.0^{\Diamond },6104.8 & %
12.99 & 6.42 & 6.45 & 1.73 & 6089.1 \\
\Xi _{b}^{\prime }(\text{MeV}) & 6226.9^{\Diamond },6235.8,6243.4,6252.3,6262.5
& 9.37 & 6.29 & 5.76 & 1.48 & 6249.1 \\
\hline\hline
\end{array}\]
\end{table}

This way of parameter relaxation via the nonlinear mass relation overcome
the difficulty in previous works\cite{KarlinerR:D15,JiaLH:19} within the scheme of the $jj$ coupling, that the masses of the baryons $\Sigma _{c}/\Xi _{c}^{\prime }$ and $\Sigma _{b}/\Xi _{b}^{\prime }$ remain undetermined due to unknown $b$: a simple comparison of the $b$-dependent mass predictions \cite{KarlinerR:D15,JiaLH:19} with experimental masses yields unreasonable values of $b$ except for the $\Sigma _{c}$. In our computation of the P-wave masses of the $\Sigma_{c}/\Xi _{c}^{\prime }$'s and $\Sigma_{b}/\Xi_{b}^{\prime }$'s, using of appropriate eigenstates, the bases of the $Jls$ coupling here, is crucial to find stable parameters which keep them within a narrow range of a few MeV around the initial inputs in Table VII. We await upcoming measurement(mass and spin-parity) by LHCb and Belle to verify the predictions in Table VIII.

Similar procedure can apply to the excited $\Omega _{b}$, giving nearly same
parameters and masses as Eqs. (\ref{a11}) and (\ref{Momi}), only that the mass of the $J=5/2$ state evolves to $M(5/2,2^{\prime})=6352.2$ MeV, $0.7$ MeV above that in Eq. (\ref{Momi}), and $c$ becomes a non-vanishing but small value of $c=1.35$, as it should be.
\medskip

\section{Spin-couplings in relativized quark model}
One may ask if the parameters of spin interaction (\ref{Hsd}) can be computed based on dynamics of quark interaction, as it is done for fine and hyperfine structures of the hydrogen-like atoms? In this section, we employ a QCD analogues of Breit-Fermi spin interaction(Appendix C), derived from the Lorentz-invariant quark-antiquark scattering amplitude at the tree level\cite{Yndurain83}, to compute the spin-couplings ($a_{1},a_{2},b, c$) and compare them with the experimentally matched ones.

Including the spin-dependent forces(Breit-Fermi spin interaction), the quasi-static potential between quark $1$ and antiquark $2$ is \cite{Yndurain83,GI:D85},
\begin{equation}
\begin{array}{r}
V^{\text{quasi-static}}=V+S+\left( \frac{V^{\prime }-S^{\prime }}{r}\right)
\mathbf{L}\cdot \left( \frac{\mathbf{S}_{1}}{2m_{d}^{2}}+\frac{\mathbf{S}_{2}%
}{2m_{Q}^{2}}\right)  \\
+\left( \frac{V^{\prime }}{r}\right) \mathbf{L}\cdot \left( \frac{\mathbf{S}%
_{1}+\mathbf{S}_{2}}{m_{1}m_{2}}\right) +\frac{1}{3m_{1}m_{2}}\left( \frac{%
V^{\prime }}{r}-V^{\prime \prime }\right) S_{12} \\
+\frac{2}{3m_{1}m_{2}}\left( \mathbf{\nabla }^{2}V\right) \mathbf{S}%
_{1}\cdot \mathbf{S}_{2},%
\end{array}
\label{Vq}
\end{equation}%
where $V$ and $S$ are the respective vector and scalar potentials, and $%
V^{\prime }$, $S^{\prime }$ and $V^{\prime \prime }$ their derivatives.

One way to relativize the instant potential $V^{\text{quasi-static}}$ is to use
the replacements\cite{GI:D85}
\begin{eqnarray}
V(r) &\rightarrow &\tilde{V}(r)=\left( \frac{m_{1}m_{2}}{E_{1}E_{2}}\right)
^{1/2+\epsilon _{V}/2}V(r)\left( \frac{m_{1}m_{2}}{E_{1}E_{2}}\right)
^{1/2+\epsilon _{V}/2},  \notag \\
S(r) &\rightarrow &\tilde{S}(r)=\left( \frac{m_{1}m_{2}}{E_{1}E_{2}}\right)
^{1/2+\epsilon _{S}/2}S(r)\left( \frac{m_{1}m_{2}}{E_{1}E_{2}}\right)
^{1/2+\epsilon _{S}/2},  \label{VSh}
\end{eqnarray}%
where the relativistic factors $m_{i}/E_{i}=m_{i}/\sqrt{m_{i}^{2}+|\mathbf{p|%
}^{2}}=$ $\sqrt{1-v_{i}^{2}}$($i=1,2$) rise from the length contraction $%
r\rightarrow r(m/E)$ when quark $i$ moves relativistically, and these
factors tend to unity in the low-momentum ($p/m_{i}\rightarrow 0$) limit of
the quarks. Here, $\epsilon _{V,S}$ are small homogeneous factors to be
determined. Applying Eqs. (\ref{VSh}) and (\ref{abc}) to a heavy quark $Q$
in color triplet($3_{c}$) and a diquark $d=qq$($q=u,d,s$) in the
anti-color triplet($\bar{3}_{c}$), four spin-couplings become then
\begin{equation}
\begin{array}{r}
a_{1}=\frac{1}{2m_{d}}\left\langle \frac{\tilde{V}^{\prime }-\tilde{S}%
^{\prime }}{m_{d}r}+\frac{2\tilde{V}^{\prime }}{m_{Q}r}\right\rangle , \\
a_{2}=\frac{1}{2m_{Q}}\left\langle \frac{\tilde{V}^{\prime }-\tilde{S}%
^{\prime }}{m_{Q}r}+\frac{2\tilde{V}^{\prime }}{m_{d}r}\right\rangle , \\
b=\frac{1}{3m_{d}m_{Q}}\left\langle \tilde{V}^{\prime }/r-\tilde{V}^{\prime
\prime }\right\rangle ,c=\left\langle \frac{2\mathbf{\nabla }^{2}\tilde{V}}{%
3m_{d}m_{Q}}\right\rangle ,%
\end{array}
\label{abc}
\end{equation}%
in which $V+S$ stands for the confining potential between $Q$ and $d$, to be
approximated by that for the heavy quarkonia $Q\bar{Q}$, and the quantum
average $\left\langle {}\right\rangle $ is made over the P-wave wavefunction
$\Psi _{Qd}$ of the $Qd$ system. For the spin interaction in Eq. (\ref{Hsd}), the relativistic treatment (\ref{VSh}) implies that the factor $1/m_{i}$ in Eq.  (\ref{abc}) becomes $1/E_{i}=(m_{i}^{2}+p^{2})^{-1/2}$, which makes sure, for example, that the hyperfine interactions of the light quark $i$ do not blow up like $1/m_{i}$ in the chiral limit($m_{i}\rightarrow 0$), but rather they should
have a finite limit determined by $\left\langle p^{-1}\right\rangle $, which
is in turn controlled by the radius of confinement.

To find P-wave wavefunction $\Psi _{Qd}$ of singly heavy baryons($Qd$) we consider the relativized quark model\cite{GI:D85} with the color-coulomb plus linear potential $V+S$(treating the spin interaction $H^{SD}$ perturbatively)
\begin{eqnarray}
H &=&\sqrt{m_{Q}^{2}+p^{2}}+\sqrt{m_{d}^{2}+p^{2}}+V(r)+S(r),  \notag \\
  V(r)&=& -k_{s}/r, S(r)=ar+C_{0}.  \label{HV}
\end{eqnarray}%
where $k_{s}=4\alpha _{s}/3$ and $\alpha_{s}$ is the strong coupling.

By introducing auxiliary fields $\mu_{d}$ and $\nu$, one can solve the model (\ref{HV}) using auxiliary field (AF) method\cite{SemayBS:D2008}(Appendix C). The inputs of $m_{d}, a$, and $\alpha_{s}$ are given in Table IX, where the values of other parameters $\xi, \zeta$ and $\epsilon_{V,S}$ for computing the spin couplings are also listed. The solved P-wave radial wavefunction of the quark-diquark system are given in FIG. 3.

\begin{table}[th]
\caption{Inputs for the model parameters and diquark masses defined in
Hamiltonian (\protect\ref{HF}) (Appendix C) for the P-wave heavy baryons listed. The heavy quark masses $m_{Q}=M_{Q}=1440$MeV($Q=c$) and $4480$MeV($Q=b$), and
light diquark masses $m_{d}$ are taken from Ref. \cite{JiaLH:19}. The confining
potential parameter $a$ is set by Regge trajectories of corresponding
baryons, taken from Ref. \cite{JiaLH:19} and Eq. (24) and Sect. IV.}\centering%
\begin{equation*}
\begin{array}{cccccccc}
\hline\hline
\text{State} & m_{d} & a\text{[GeV}^{2}\text{]} & \alpha_{s} & \xi \text{%
(GeV}^{-1}\text{)} & \zeta \text{(GeV}^{-2}\text{)} & \epsilon _{V} &
\epsilon _{S} \\ \hline
\left\{
\begin{array}{r}
\Omega _{c} \\
\Omega _{b}%
\end{array}%
\right. & 991 &
\begin{array}{r}
0.316 \\
0.318%
\end{array}
&
\begin{array}{r}
0.561 \\
0.543%
\end{array}
& 0.818 & 0.11 & -0.10 & 2.65 \\
\left\{
\begin{array}{r}
\Xi _{c}^{\prime } \\
\Xi _{b}^{\prime }%
\end{array}%
\right. & 872 &
\begin{array}{r}
0.255 \\
0.307%
\end{array}
&
\begin{array}{r}
0.590 \\
0.545%
\end{array}
& 0.820 & 0.12 & -0.06 & 2.20 \\
\left\{
\begin{array}{r}
\Sigma _{c} \\
\Sigma _{b}%
\end{array}%
\right. & 745 &
\begin{array}{r}
0.212 \\
0.246%
\end{array}
&
\begin{array}{r}
0.595 \\
0.549%
\end{array}
& 0.850 & 0.12 & -0.05 & 1.80 \\ \hline\hline
\end{array}%
\end{equation*}%
\end{table}

\begin{figure}[!h]
\centering
\begin{tabular}{c}
\includegraphics[width=0.84\textwidth]{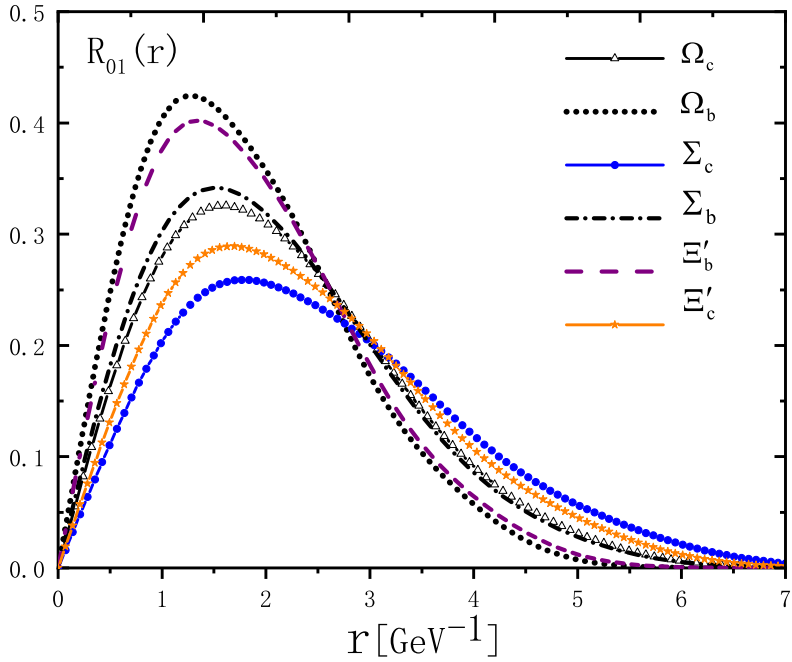}
\end{tabular}
\caption{(Color online). The P-wave radial wavefunction $R_{nL}(r)$ with $n=0$ and $L=1$, numerically solved from Eq. (\ref{PQd}), with the inputs in Table IX. } \label{figRadial3}
\end{figure}

Upon using (\ref{abc}) to compute the spin couplings, we take into
account non-point nature of the diquark $d$ via adding a form factor $%
F(r)\equiv 1-e^{-\xi r-\zeta r^{2}}$ \cite{EFG:D11} in the short-range potential $V$ so
that
\begin{equation}
V\rightarrow VF=-\frac{k_{s}}{r}F(r).  \label{run}
\end{equation}
Putting the improved $V$ in Eq. (\ref{run}) into Eq. (\ref{abc}), in which
the relativized potentials become now, by Eq. (\ref{VSh}), $%
\tilde{V}=(m_{d}m_{Q}/(E_{d}^{H}E_{Q}^{H}))^{1+\epsilon _{V}}V$ and $\tilde{S%
}=(m_{d}m_{Q}/(E_{d}^{N}E_{Q}^{N}))^{1+\epsilon _{S}}S$ with $S=ar+C_{0}$
and
\begin{equation}
E_{i}^{H,N}=[m_{i}^{2}+\langle p^{2}\rangle _{H,N}]^{1/2},  \label{ENH}
\end{equation}%
one finds
\begin{equation}
\begin{array}{r}
a_{1}=\frac{1}{2m_{d}^{2}}\left\langle \left( 1+\frac{2m_{d}}{m_{Q}}\right)
\left( \frac{m_{d}m_{Q}}{E_{d}^{H}E_{Q}^{H}}\right) ^{1+\epsilon
_{V}}k_{s}\left( \frac{F}{r^{3}}-\frac{F^{\prime }}{r^{2}}\right) -\left(
\frac{m_{d}m_{Q}}{E_{d}^{N}E_{Q}^{N}}\right) ^{1+\epsilon _{S}}\frac{a}{r}%
\right\rangle , \\
a_{2}=\frac{1}{m_{d}m_{Q}}\left\langle \left( 1+\frac{m_{d}}{2m_{Q}}\right)
\left( \frac{m_{d}m_{Q}}{E_{d}^{H}E_{Q}^{H}}\right) ^{1+\epsilon
_{V}}k_{s}\left( \frac{F}{r^{3}}-\frac{F^{\prime }}{r^{2}}\right) -\left(
\frac{m_{d}m_{Q}}{E_{d}^{N}E_{Q}^{N}}\right) ^{1+\epsilon _{S}}\frac{m_{d}a}{%
2m_{Q}r}\right\rangle , \\
b=\frac{k_{s}}{3m_{d}m_{Q}}\left( \frac{m_{d}m_{Q}}{E_{d}^{H}E_{Q}^{H}}%
\right) ^{1+\epsilon _{V}}\left\langle \frac{3F}{r^{3}}-\frac{3F^{\prime }}{%
r^{2}}+\frac{F^{\prime \prime }}{r}\right\rangle , \\
c=\frac{2k_{s}}{3m_{d}m_{Q}}\left( \frac{m_{d}m_{Q}}{E_{d}^{H}E_{Q}^{H}}%
\right) ^{1+\epsilon _{V}}\left[ 4\pi |R_{nL}^{H}(0)|^{2}-\int drrF^{\prime
\prime }|R_{nL}^{H}(r)|^{2}\right] ,%
\end{array}
\label{aabc}
\end{equation}%
where we have used $V^{\prime }=k_{s}(F/r^{2}-F^{\prime }/r)$, $V^{\prime
\prime }=k_{s}(2F^{\prime }/r^{2}-2F/r^{3}-F^{\prime \prime }/r)$, $%
S^{\prime }=a$, and the Laplace relation $\mathbf{\nabla }^{2}(F/r)=-4\pi
\delta ^{3}(\mathbf{r})+F^{\prime \prime }/r$. Here, the average is taken
over $R_{0L}(r)$, and for $n_{r}=0$,
\begin{eqnarray}
E_{d}^{H} &=&\mu _{dH},E_{Q}^{H}=\left[ m_{Q}^{2}+\frac{(k_{s}\mu _{H})^{2}}{%
(L+1)^{2}}\right] ^{1/2},  \notag \\
E_{i}^{N} &=&\left[ m_{i}^{2}+a\left( L+\frac{3}{2}\right) \sqrt{\mu /\nu }%
\right] ^{1/2},i=Q,d,  \label{EHN}
\end{eqnarray}%
where the average $\langle p^{2}\rangle _{N}=\alpha ^{2}(L+3/2)=(a^{2}\mu
/v)^{1/2}(L+3/2)$ in the HO wavefunction is used again.

The AF fields ($\mu, \nu $) are solvable from the nonlinear equation (\ref{ITer}) (Appendix C) given the inputs of the masses $m_{Q}=M_{Q}$ and $m_{d}$, and the linear potential parameter $a$ extracted from Ref. \cite{JiaLH:19}, as listed in Table IX, by which  Eq. (\ref{EHN}) gives the inverse HO length $%
\alpha $, and the relativistic factors $m_{i}/E_{i}^{H,N}$. Notice that two
derivatives $F^{\prime }=(2\zeta r+\xi )e^{-\xi r-\zeta r^{2}},F^{\prime
\prime }=[2\zeta -(2\zeta r+\xi )^{2}]e^{-\xi r-\zeta r^{2}}$, one can then
compute the spin-couplings in Eqs. (\ref{aabc}) via quantum averaging with $%
R_{nL}(r)$ solved from Eq. (\ref{PQd})(Appendix C). The results are shown in Table X, which agree reasonably with the parameters ($%
a_{1},a_{2},b$) given in Eq. (12) and Eq. (25) as well as that in Table VIII. One sees from Table X that our prediction for the spin couplings $a_{2}$ is
slightly overestimated compared to that matched ones, about $4-8$MeV, and
the predictions for $c$ is anomaly larger than the matched ones, about $7-16$%
MeV.

\begin{table}[h]
\caption{Spin coupling parameters $a_{1}$, $a_{2}$, $b$ and $c$ calculated by Eq. (\protect\ref{aabc}), the AF fields ($\protect\mu ,\protect\nu $) solved from
Eq. (\protect\ref{ITer}) (Appendix C) and other related parameters for the P-wave
states. All in MeV except for $a$. }
\label{Table 5}$%
\begin{array}{ccccccccccc}
\hline\hline
\text{State} & \mu _{d} & \mu & \nu & \mu _{dH} & \mu _{H} & a_{B}\text{[GeV}%
^{-1}\text{]} & a_{1} & a_{2} & b & c \\ \hline
\Omega _{c} & 1291 & 681 & 909 & 1016 & 596 & 2.248 & 28.52 & 27.03 & 15.32
& 20.73 \\
\Omega _{b} & 1379 & 1054 & 789 & 1034 & 840 & 1.701 & 10.30 & 10.25 & 5.61
& 9.26 \\
\Xi _{c}^{\prime } & 1153 & 640 & 805 & 900 & 554 & 2.256 & 30.15 & 27.98 &
16.54 & 20.35 \\
\Xi _{b}^{\prime } & 1273 & 992 & 787 & 912 & 757 & 1.882 & 11.42 & 11.06 &
6.14 & 9.82 \\
\Sigma _{c} & 1017 & 596 & 728 & 777 & 504 & 2.283 & 35.46 & 30.96 & 19.10 &
20.44 \\
\Sigma _{b} & 1117 & 894 & 703 & 781 & 665 & 2.130 & 13.93 & 11.15 & 6.61 &
8.93 \\ \hline\hline
\end{array}%
$%
\end{table}

One can list the resulted ratios of the spin couplings $%
r_{1,2}=a_{1,2}(bqq)/a_{1,2}(cqq)$ and $r_{b}=b(bqq)/b(cqq)$ between the bottom
and charmed baryons and compare them to corresponding ratios (denoted as $[.]_{Match}$) obtained from Eqs.(12),(25) and from Table VIII:%
\begin{equation}
\begin{array}{cccc}
\text{ratio:} & r_{1} & r_{2} & r_{b} \\
\Omega _{b}/\Omega _{c}: & 0.361 & 0.379 & 0.366 \\
\Xi _{b}^{\prime }/\Xi _{c}^{\prime }: & 0.379 & 0.395 & 0.370 \\
\Sigma _{b}/\Sigma _{c}: & 0.393 & 0.360 & 0.346%
\end{array}%
\begin{array}{ccc}
\lbrack r_{1}]_{Match} & [r_{2}]_{Match} & [r_{b}]_{Match} \\
0.33 & 0.16 & 0.56, \\
0.28 & 0.31 & 0.35, \\
0.33 & 0.29 & 0.31,%
\end{array}
\label{rat}
\end{equation}%
One sees that the simple mass scaling law($\sim
M_{b}/M_{c}=1440/4480=\allowbreak 0.321$) holds approximately for $a_{2}$ and $b$. In the case of $a_{1}$, the scaling law $a_{1}(bqq)=a_{1}(cqq)$ \cite{KarlinerR:D15} is broken somehow as $r_{1}\sim 1/3$ is less than unity. Further, the matched value $r_{2}=0.16$ for the baryons $\Omega _{b}$ vs. $\Omega _{c}$ is somehow smaller than that expected by $M_{b}/M_{c}$ and also than our computed value $r_{2}=0.379$.

From Table XI, which shows the computed values of the
relevant factors and the averages in Eq. (\ref{aabc}), one sees that the minus term$\sim $ $%
-\left\langle m_{d}/m_{Q}\ a/r\right\rangle $ in Eq. (\ref{aabc}) is
notably suppressed and this gives relatively larger predictions for $a_{2}$. We remark three features of our computations:

(i) As a leading approximation, the Breit-Fermi-like interaction (\ref{Vq})
have not considered nonperturbative corrections that may enter at
long-distance regime. Assuming the spin interaction dominates at short
distance, one expects that the interaction (\ref{Vq}) includes the main
spin-spin and spin-orbital forces via adding the scalar confining potential $%
S\sim r$ to the vector potential $V$, and such a nonperturbative corrections
to Eq. (\ref{Vq}) is relatively small.

(ii) Static approximation of quark-diquark potential $V+S$, which is valid
in principle only for heavy quarkonia($Q\bar{Q}$), has ignored rotation of
QCD string when $L\neq 0$ \cite{LaCourseO:D89,KalashNef:D00,AgaevAS:D17}. In orbitally-excited state, the
rotation of QCD string can change the static quark-diquark potential $V+S$
in Eq. (\ref{HV}), as shown in Refs. \cite{LaCourseO:D89,KalashNef:D00}.
For instance, this corrects $L$-dependence of the Hamiltonian, $%
[m_{d}^{2}+p_{r}^{2}+L^{2}/r^{2}]^{1/2}\sim L$ into a Regge-like one, $H\sim
\sqrt{\pi aL} $ due to addition of the orbital angular moment of the
string \cite{LaCourseO:D89}.

(iii) A simple exponential form factor $F(r)$ may not be enough in the
short-range to describe the three-body quark systems of $Q(qq)$. Three-body
interaction \cite{CI:D86,SemayBS:D08} inside baryons may make $F(r)$ quite nontrivial as the
effective running of the color-charge of a non-point diquark $d$ in
color-antitriplet ($\bar{3}_{c}$) seen by $Q$ in color-triplet ($3_{c}$)
becomes highly distance-dependent when $d$ is close to $Q$. See Ref.
\cite{CI:D86}(Eq. (18)) for other possible forms of the form factor, like $%
Erf[c_{0}r] $ or its linear combinations. For simplicity, we used the $%
R_{nL}^{H}(r)$ in Eq. (\ref{RH}) (Appendix C) to estimate $c$ in Eq. (\ref{aabc}) and
this may overestimate values for $c$ govern by integral of $-F^{\prime
\prime }|R_{nL}^{H}(r)|^{2}$ at short-range. Note that $R_{nL}^{H}(0)\equiv 0$
for the P-waves.

\begin{table}[h]
\caption{Values of the relevant factors (dimensionless) and the computed averages(in GeV$^{-3}$) for spin coupling $a_{1,2}$ and $b$ in Eq. (\protect
\ref{aabc}).}
\label{Table 3}%
\begin{equation*}
\begin{array}{cccccccc}
\hline\hline
\text{State} & \frac{m_{d}}{m_{Q}} & \left( \frac{m_{d}m_{Q}}{%
E_{d}^{H}E_{Q}^{H}}\right) ^{1+\epsilon _{V}} & \left( \frac{m_{d}m_{Q}}{%
E_{d}^{N}E_{Q}^{N}}\right) ^{1+\epsilon _{S}} & \left\langle \frac{m_{d}\ a}{%
2m_{Q\ }r}\right\rangle & \left\langle \frac{F}{r^{3}}-\frac{F^{\prime }}{%
r^{2}}\right\rangle & \left\langle \frac{a}{r}\right\rangle & \left\langle
\frac{3F}{r^{3}}-\frac{3F^{\prime }}{r^{2}}+\frac{F^{\prime \prime }}{r}%
\right\rangle \\ \hline
\Omega _{c} & 0.688 & 0.884 & 0.227 & 0.0472 & 0.056 & 0.137 & 0.0994 \\
\Omega _{b} & 0.220 & 0.866 & 0.276 & 0.0182 & 0.075 & 0.164 & 0.123 \\
\Xi _{c}^{\prime } & 0.606 & 0.864 & 0.278 & 0.0309 & 0.049 & 0.102 & 0.090
\\
\Xi _{b}^{\prime } & 0.195 & 0.861 & 0.278 & 0.0149 & 0.071 & 0.153 & 0.119
\\
\Sigma _{c} & 0.518 & 0.833 & 0.313 & 0.0204 & 0.044 & 0.079 & 0.085 \\
\Sigma _{b} & 0.167 & 0.849 & 0.306 & 0.0092 & 0.062 & 0.110 & 0.111 \\
\hline\hline
\end{array}%
\end{equation*}%
\end{table}

In addition, Eq. (\ref{aabc}) seems to imply the scaling laws $a_{1}\sim \mathcal{O}%
(m_{Q}^{0})$, $a_{2}$ $\sim 1/m_{Q}\sim b$ (see Eq. (24) in Ref. \cite{KarlinerR:D15})), as
heavy quark symmetry expected. However, the small variation of the averages within the baryons, manly in $1+2m_{d}/m_{Q}$ and $k_{s}=4\alpha _{s}/3$ along
with the size($\sim a_{B}$) of baryons, is enlarged by $m_{d}^{2}$ in
denominator for $a_{1}$ in Eq. (\ref{aabc}) and this makes $a_{1}$ changes
notably between the bottom and charmed sectors. This kind of enlargement
does not happen for other spin couplings due to their overall suppression factor $1/m_{d}m_{Q}$. Note also that the factor $1+m_{d}/2m_{Q}$ for $a_{2}$ changes slightly with heavy flavor $Q(=c,b)$, compared to the variations of $%
1+2m_{d}/m_{Q}$:
\begin{eqnarray*}
\left( 1+\frac{m_{d}}{2m_{Q}}\right) _{c,b}
&:&\{1.34,1.11\}_{ssQ},\{1.30,1.09\}_{snQ},\{1.26,1.08\}_{nnQ}, \\
\left( 1+\frac{2m_{d}}{m_{Q}}\right) _{c,b}
&:&\{2.37,1.44\}_{ssQ},\{2.21,1.39\}_{snQ},\{2.04,1.33\}_{nnQ},
\end{eqnarray*}%
As seen in Tables X and XI, the scaling law for the spin couplings breaks moderately, especially for $a_{1}$, due to the injured heavy quark symmetry indicated by Eq. (\ref{aabc}).

\section{Conclusions and Discussions}
Exploring and building data set of the mean-spin mass $\bar{M}$ and the spin coupling strengths($a_{1},a_{2}, b, c$) is important as it can be a basis to explore how quarks interact within hadrons. Although baryon is a three-body system with subtle short-range interactions, it is useful to employ a simple heavy quark-diquark picture for the singly heavy baryons to understand the  measured mass data of excited heavy-baryons. In this work, we have proposed a new scheme of state classification, the $Jls$ mixing coupling, to unify analysis of masses and inner-structures of the excited baryons $\Omega_{Q}$ and other less-strange heavy baryons with spin-1 light diquarks. Firstly, we interpret the five $\Omega _{c}$ and four $\Omega _{b}$ excited states reported by LHCb to be the relative $P$-wave excitations between spin-1 $ss$ diquark and the heavy quark, the charm quark $c$ for the $\Omega _{c}$ and the bottom quark $b$ for the $\Omega _{b}$, respectively. Further analysis via Regge phenomenology for the low-lying excited states excludes the possibility that some of higher states of the resonances $\Omega _{c}$ and $\Omega _{b}$ are the $2S$ states since the $2S$ candidates are all too heavy to accommodate both of the resonances. Secondly, we applied our formalism based on the $Jls$ mixing coupling to update the earlier computation of all P-wave masses of the less strange baryons $\Sigma _{c}/\Xi _{c}^{\prime}$ and $\Sigma _{b}/\Xi _{b}^{\prime}$, for which only single measured mass available for each of these baryons. The masses of all spin-partners of the baryons $\Sigma _{c}(2800)/\Xi_{c}^{\prime }(2942)$ and $\Sigma _{b}(6097)$/$\Xi _{b}^{\prime }(6227)$ are then predicted. Finally, we demonstrated that the matched spin-couplings of the singly heavy baryons considered can be understood in a relativized potential quark model, based on which the scaling law for their spin coupling parameters are discussed.

With the help of the relation (\ref{MM}), we find that four
measured masses for the excited $\Omega _{b}$'s are enough to make a quantitative
prediction $M(5/2^{-})=6352$ MeV for one unseen state with $J^{P}=5/2^{-}$. The missing of this $5/2^{-}$ state in LHCb experiment can be due to the degeneracy with the nearby state $|3/2^{-},2^{\prime }\rangle $ at $6350$ MeV. The same $J^{P}=5/2^{-}$ state is predicted as the unseen state in a recent assignment \cite{KarlinerR:Ax20}, with two higher mass ranges predicted, about $10$ MeV away or more.

One very recent prediction for the odd-parity $\Omega_{b}$'s is that by the LS coupling\cite{Mutuk:A20}:
\begin{eqnarray}
M(1/2^{-})=6314,6330 \text{ MeV},\notag \\
M(3/2^{-})=6339,6342 \text{ MeV},M(5/2^{-})=6352 \text{ MeV}, \label{LSP}
\end{eqnarray}
for which the explanation of near degeneracy applies only to two $J=3/2$ states and assuming one of them to be unseen is favored. In the viewpoint of this work, with given mixing weights in Eq. (\ref{In0}), two LS bases $|^{2}P_{3/2}\rangle $ and $|^{4}P_{3/2}\rangle $ are not eigenstates of the Hamiltonian (\ref{Hsd}) for $J=3/2$ and tend to mix deeply to form the true $Jls$ eigenstates $|3/2,1^{\prime }\rangle$ and $|3/2,2^{\prime }\rangle $, with respective masses $6337.8$ MeV and $6350.0$ MeV in Eq. (\ref{Momi}) both matching the measured masses($6339.7$ MeV and $6349.9$ MeV) nicely. Our mass predictions in Eq. (\ref{Momi}) for other states($J=3/2, 5/2$) should agree with that by the LS coupling because the bases for the LS and Jls couplings are same($J=5/2$) or roughly same($J=3/2$), as two predictions(Eq. (\ref{LSP}) v.s. Eq. (\ref{Momi})) indicated. This favors the state at $M(5/2^{-})=6352$ MeV to be the unseen state.

Our computation via a simple heavy quark-diquark picture indicates that the measured masses of the excited heavy-baryons considered can be understood via a relativized potential quark model and resulted Breit-Fermi spin-orbit and tensor forces, provided that a proper relativizing is made to these spin interactions(forces). By the way, we confirm based on the relativized expression of the Breit-Fermi formula (\ref{aabc}) for the spin coupling parameters that a mass scaling law\cite{KarlinerR:D15} for them holds approximately for all baryons considered. Being at the order of $\mathcal{O}(1)$, the scaling relation $a_{1}(bqq)=a_{1}(cqq)$ for $a_{1}$ is broken most, in contrast with other parameters at the order of $\mathcal{O}(1/m_{Q})$. Due to involved QCD dynamics of the heavy flavor baryons(see \cite{EFG:D11,Manohar:D07} for instance), our analysis for the spin-couplings is of quite approximative, with further quantitative corrections remained to be explored.

\medskip

\textbf{ACKNOWLEDGMENTS}

D. J thanks Xiang Liu and A. Hosaka for discussions. D. J. is supported by
the National Natural Science Foundation of China under the no. 11565023 and
the Feitian Distinguished Professor Program of Gansu (2014-2016). C.P is
supported by the National Natural Science Foundation of China under Grants
No. 11965016, and by the projects funded by Science and Technology
Department of Qinghai Province (No. 2018-ZJ-971Q and No. 2019-ZJ-A10), the
Key Laboratory of IoT of Qinghai under grant (No. 2020-ZJ-Y16).

\medskip
\section*{Appendix A}
\setcounter{equation}{0}
\renewcommand{\theequation}{A\arabic{equation}}
\medskip

For a heavy quark-diquark system $Qd$, the bases of the mixing $Jls$
coupling are the eigenfunctions of the mass operator $H^{mix}=\mathbf{L}%
\cdot \mathbf{S}_{d}+\epsilon \mathbf{L}\cdot \mathbf{S}_{Q}+b_{1}S_{12}$,
with $b_{1}\equiv b/a_{1}$. They can be obtained by diagonalizing $H^{mix}$,
that is, diagonalizing the $2\times 2$ matrices in Eqs. (\ref{M2})-(\ref{M3}%
) separately in the subspaces of $J=1/2$, $J=3/2$. When $J=1/2$, the matrix

\begin{equation}\label{A1}
H_{J=1/2}=\left[
\begin{array}{cc}
\frac{1}{3}(\epsilon -4) & \frac{\sqrt{2}}{3}(\epsilon -1)+\frac{b_{1}}{%
\sqrt{2}} \\
\frac{\sqrt{2}}{3}(\epsilon -1)+\frac{b_{1}}{\sqrt{2}} & -\frac{5}{3}(1+%
\frac{\epsilon }{2})-b_{1}%
\end{array}%
\right]
\end{equation}%
can be diagonalized to give $M_{1/2}=\langle H_{J=1/2}\rangle $. Thus, adding a perturbative correction $\sim c_{1}M_{c}$ to $M_{1/2}$, with
\begin{equation}\label{A4}
M_{c}=\left[
\begin{array}{cc}
-1 & 0 \\
0 & \frac{1}{2}%
\end{array}%
\right] ,
\end{equation}%
leads to the mass shifts in the $J=1/2$ subspace:
\begin{eqnarray}\label{A5}
\Delta M(1/2,0^{\prime })/a_{1} &=&-\frac{3}{2}-\frac{1}{2}b_{1}-\frac{1}{4}%
\epsilon -\frac{1}{4}\sqrt{\Delta _{1}(\epsilon ,b_{1})}+c_{1}\Delta
_{3}^{+},  \notag \\
\Delta M(1/2,1^{\prime })/a_{1} &=&-\frac{3}{2}-\frac{1}{2}b_{1}-\frac{1}{4}%
\epsilon +\frac{1}{4}\sqrt{\Delta _{1}(\epsilon ,b_{1})}+c_{1}\Delta
_{3}^{-},
\end{eqnarray}%
with the functions $\Delta _{1}(\epsilon ,x),\Delta _{3}^{\pm }(\epsilon ,x)$
defined in Eqs. (\ref{del12}) and (\ref{del3}).

In the $J=3/2$ subspace, a similar diagonalization of $H_{J=3/2}$ which gives $M_{3/2}$, leads to, by adding a perturbative correction $\sim c_{1}M_{c}$,
\begin{eqnarray}\label{A8}
\Delta M(3/2,1^{\prime })/a_{1} &=&\frac{2}{5}b_{1}-\frac{1}{4}\epsilon -%
\sqrt{\Delta _{2}(\epsilon ,b_{1})}+c_{1}\Delta _{4}^{+},  \notag \\
\Delta M(3/2,2^{\prime })/a_{1} &=&\frac{2}{5}b_{1}-\frac{1}{4}\epsilon +%
\sqrt{\Delta _{2}(\epsilon ,b_{1})}+c_{1}\Delta _{4}^{-},
\end{eqnarray}%
with the functions $\Delta _{2}(\epsilon ,x)$ and $\Delta _{4}^{\pm
}(\epsilon ,x)$ defined in Eqs. (\ref{del12}) and (\ref{del4}). In the $J=5/2$
subspace, $\Delta M/a_{1}$ is simply $\Delta M(5/2,2^{\prime
})/a_{1}=1-b_{1}/5+\epsilon /2+c_{1}/2$. This, combined with Eq. (A5) and
(A8), proves Eq.(\ref{MM}). A recent Ref. \cite{Ali:jh2019} gives an alternative expression similar to the mass-shifts (\ref{MM}).

\medskip

\section*{Appendix B}

\medskip
\setcounter{equation}{0}
\renewcommand{\theequation}{B\arabic{equation}}
The coefficients of Eq. (\ref{VV}) form normalized vectors of the
eigenstates of the operator $H^{mix}$. In the subspaces of $J=1/2$, $H^{mix}$
becomes the matrix $H_{J=1/2}$ in Eq. (A1) with the eigenstates given in Eq.
(A3), which are $\mathbf{v}_{1}=\{-0.208,1\}$ and $\mathbf{v}%
_{2}=\left\{ 4.804,1\right\} $ using Eq. (\ref{best}). Upon normalization, the eigenstates give rise to the coefficients in the first and second lines of Eq. (\ref{VV}). Similar computation leads to the coefficients
in the third and fourth lines of Eq. (\ref{VV}).  

To write the mixing bases $|J,j^{\prime }\rangle $ in terms of the $jj$
bases $|J,j\rangle $, one writes firstly the $P$-wave $|J,j\rangle $ in
terms of the $L-S$ eigenstates $|1^{2S+1}P_{J}\rangle $, as given in Eq.(\ref%
{JJS}). Its inverse is%
\begin{eqnarray}\label{B1}
|1^{2}P_{1/2}\rangle  &=&\sqrt{\frac{1}{3}}|1/2,j=0\rangle +\sqrt{\frac{2}{3}%
}|1/2,j=1\rangle ,  \notag \\
|1^{4}P_{1/2}\rangle  &=&\sqrt{\frac{2}{3}}|1/2,j=0\rangle -\sqrt{\frac{1}{3}%
}|1/2,j=1\rangle .
\end{eqnarray}
Putting it into the first two lines of Eq. (\ref{VV}) gives the first and
second lines of Eq.(\ref{Jjj}). Similarly, one can obtain the third and fourth lines of Eq.(\ref{Jjj}) with the help of
\begin{eqnarray}\label{B2}
|J &=&3/2,j=1\rangle =\sqrt{\frac{1}{6}}|1^{2}P_{3/2}\rangle +\sqrt{\frac{5}{%
6}}|1^{4}P_{3/2}\rangle ,  \notag \\
|J &=&3/2,j=2\rangle =\sqrt{\frac{5}{6}}|1^{2}P_{3/2}\rangle -\sqrt{\frac{1}{%
6}}|1^{4}P_{3/2}\rangle ,
\end{eqnarray}%
and its inverse transformation.

\section*{Appendix C}
\medskip
\setcounter{equation}{0}
\renewcommand{\theequation}{C\arabic{equation}}
Using auxiliary field (AF) method\cite{SemayBS:D2008}, which re-expresses nonlinear operator $%
\sqrt{O}=\min_{\lambda >0}[O/(2\lambda )+\lambda /2]$ in terms of the AF
field $\lambda $, for which the minimization is achieved when $\lambda =$ $%
\sqrt{O}$, one can rewrite the model (\ref{HV}) as
\begin{eqnarray}
H &=&m_{Q}+\frac{p^{2}}{2m_{Q}}+\frac{m_{d}^{2}+p^{2}}{2\mu _{d}}+\frac{%
(ar)^{2}}{2\nu }+\frac{\mu _{d}+\nu }{2}-\frac{k_{s}}{r}+C_{0},  \label{HF}
\\
\mu _{d} &=&\left\langle \sqrt{m_{d}^{2}+p^{2}}\right\rangle ,\nu
=a\left\langle r\right\rangle ,  \label{mn}
\end{eqnarray}%
where $p=|\mathbf{p}|$ is the 3-momentum of the heavy quark $Q$ and the
light diquark $d$, and the minimization conditions (\ref{mn}) for the
auxiliary fields ($\mu _{d},\nu $) are assumed to hold at semi-classical
level. In a sense, the AF method linearizes the Hamiltonian $H$ in (\ref{HV}%
) in a manner similar to mean-field method for many-body systems. Next, we
solve the P-wave radial wavefunction $R_{nL}$ of the $Qd$ system defined by
Eq. (\ref{HV}) numerically in two steps.

In the first step, we find two self-consistent AF fields ($\mu _{d},\nu $)
via solving Eq. (\ref{mn}). For this, we ignore the (color) Coulomb term $%
-k_{s}/r$ to write the Hamiltonian (\ref{HF}) as that of three dimensional
harmonic oscillator(HO) with reduced mass $\mu \equiv \mu _{d}/(1+\mu
_{d}/m_{Q})$, the HO frequency $\omega =a/\sqrt{\mu \nu }$, the inverse HO
length $\alpha =\sqrt{\mu \omega }=(a^{2}\mu /v)^{1/4}$. As such, the
wavefunction $\Psi _{Qd}$ is given by that of the HO, $\psi
^{N}=R_{n_{r}L}^{N}(r)Y_{Lm}(\theta ,\varphi )$ with quantum number $%
N=2n_{r}+L$, where $Y_{Lm}\ $is the spherical function and
\begin{equation}
R_{n_{r}L}^{N}(r)=\alpha ^{3/2}\left[ \frac{2n_{r}!}{\Gamma (n_{r}+L+3/2)}%
\right] ^{1/2}(\alpha r)^{L}L_{n_{r}}^{L+1/2}(\alpha ^{2}r^{2})e^{-(\alpha
r)^{2}/2},  \label{RN}
\end{equation}%
with $L_{n_{r}}^{2L+1}$ the associated Laguerre polynomial. For the orbital
states with $n_{r}=0$ we address, one finds from Eq. (\ref{mn}),
\begin{eqnarray*}
\mu _{d} &\simeq &\sqrt{m_{d}^{2}+\left\langle p^{2}\right\rangle _{N}}%
=[m_{d}^{2}+\alpha ^{2}(L+3/2)]^{1/2}, \\
\nu &\simeq &a\sqrt{\left\langle r^{2}\right\rangle _{N}}=a\left[ \frac{L+3/2%
}{\mu \omega }\right] ^{1/2},
\end{eqnarray*}%
where the HO-state averages of $p^{2}$ and $r^{2}$ are used. This leads to
two nonlinear equations%
\begin{eqnarray}
\mu &\equiv &\frac{[m_{d}^{2}+a(L+3/2)(\mu /\nu )^{1/2}]^{1/2}}{%
1+[m_{d}^{2}+a(L+3/2)(\mu /\nu )^{1/2}]^{1/2}/m_{Q}},  \notag \\
\nu &=&\sqrt{a(L+3/2)}(\nu /\mu )^{1/4},  \label{ITer}
\end{eqnarray}%
which can be solved by numerical iteration. The solution for ($\mu ,\nu $)
and $\mu _{d}\equiv \mu /(1-\mu /m_{Q})$ are shown in Table 1(Table IX in
revision).

The solution of the wavefunction $\Psi_{Qd}$ can be solved from the Hamiltonian (\ref{HF}) of the two-body $Qd$, which is
\begin{eqnarray}
H &=&\frac{p^{2}}{2\mu }+\frac{1}{2}m\omega ^{2}r^{2}-\frac{k_{s}}{r}+m_{Q}+%
\frac{m_{d}^{2}}{2\mu _{d}}+\frac{\mu _{d}+\nu }{2}+C_{0},  \label{HO} \\
\mu  &=&\mu _{d}/\left[ 1+\frac{\mu _{d}}{m_{Q}}\right] ,\omega =\frac{a}{%
\sqrt{\mu \nu }}.  \label{HOO}
\end{eqnarray}

In the long-range where the color-Coulomb term $-k_{s}/r$ is ignorable, the radial eigenfunctions of the Hamiltonian (\ref{HO}) can be given by that of the HO wavefunction,
\begin{equation}
R_{n_{r}L}^{N}(r)=\alpha ^{3/2}\left[ \frac{2n_{r}!}{\Gamma (n_{r}+L+3/2)}\right]
^{1/2}(\alpha r)^{L}L_{n_{r}}^{L+1/2}(\alpha ^{2}r^{2})e^{-(\alpha r)^{2}/2},
\label{RN}
\end{equation}%
with $L_{n_{r}}^{2L+1}$ the associated Laguerre polynomial.

In the second step, we use the obtained ($\mu ,\nu $) to numerically solve
the eigenstate equation for the singly heavy baryons $\Omega _{Q}$, $\Xi
_{Q}^{\prime }$ and $\Sigma _{Q}$($Q=c,b$)
\begin{equation}
\left[ -\frac{1}{2\mu r^{2}}\frac{d}{dr}\left( r^{2}\frac{d}{dr}\right) +%
\frac{L(L+1)}{2\mu r^{2}}+\frac{a^{2}}{2\nu }r^{2}-\frac{k_{s}}{r}\right]
R_{nL}=ER_{nL},  \label{PQd}
\end{equation}%
where $R_{nL}(r)$ are the radial wavefunctions of these systems. The results
for the P-wave solutions $R_{nL}(r)$ are shown in FIG. 3.

It is useful to write the radial wavefunction $R_{n_{r}L}^{H}(r)$ in the
small-$r$ range, which is given by solution to the eigen-equation of the
Hamiltonian (\ref{HF}) with the HO potential($\sim r^{2}$) ignored. It is
the solution to the hydrogen-like atoms with the effective Bohr radius $%
a_{B}=1/(\mu _{H}k_{s})$ and the reduced mass $\mu _{H}=\mu _{dH}/(1+\mu
_{dH}/m_{Q})$ with $\mu _{dH}=\left\langle \sqrt{m_{d}^{2}+p^{2}}%
\right\rangle _{H}$ averaged in the hydrogen-like wavefunction $\psi
^{H}=R_{n_{r}L}^{H}(r)Y_{Lm}(\theta ,\varphi )$, with
\begin{equation}
R_{n_{r}L}^{H}(r)=\left\{ \left( \frac{2}{na_{B}}\right) ^{3}\frac{n_{r}!}{%
2n[(n+L)!]^{3}}\right\} ^{1/2}e^{-r/(na_{B})}\left( \frac{2r}{na_{B}}\right)
^{L}L_{n_{r}}^{2L+1}\left( \frac{2r}{na_{B}}\right) ,  \label{RH}
\end{equation}%
with $n=n_{r}+L+1$. Using averaging in the hydrogen-like wavefunction $\psi
^{H}$ with $n_{r}=0$, $\langle p^{2}\rangle _{H}=k_{s}\mu
_{H}/[a_{B}(L+1)^{2}]=k_{s}^{2}\mu _{H}^{2}/(L+1)^{2}$, one finds that the
AF parameter $\mu _{dH}\simeq $ $\sqrt{m_{d}^{2}+\langle p^{2}\rangle _{H}}$
is subjected to the self-consistent equation,
\begin{equation}
\mu _{dH}=\left[ m_{d}^{2}+\frac{1}{(L+1)^{2}}\frac{(k_{s}\mu _{dH})^{2}}{%
(1+\mu _{dH}/m_{Q})^{2}}\right] ^{1/2}.  \label{mdH}
\end{equation}%
The numerical results for $\mu _{dH}\ $and thereby-resulted $\mu _{H}$ are
shown in Table V also.

In the short-range where the term $-k_{s}/r$ is dominate over the HO potential, the radial solution of wavefunction to the Hamiltonian (\ref{HO}) is that of the hydrogen-like:
\begin{equation}
R_{n_{r}L}^{H}(r)=\left\{ \left( \frac{2}{na_{B}}\right) ^{3}\frac{n_{r}!}{%
2n[(n+L+1)!]^{3}}\right\} ^{1/2}e^{-r/(na_{B})}\left( \frac{2r}{na_{B}}%
\right) ^{L}L_{n_{r}}^{2L+1}\left( \frac{2r}{na_{B}}\right) ,  \label{RH}
\end{equation}%
with $n=n_{r}+L+1$.

\end{document}